\documentclass[useAMS,usenatbib,twocolumn]{mn2e}
\usepackage{epsfig}
\usepackage{amsmath}
\newcommand{\be}{\begin{equation}}
\newcommand{\beq}{\begin{equation}}
\newcommand{\ba}{\begin{eqnarray}}
\newcommand{\ee}{\end{equation}}
\newcommand{\eeq}{\end{equation}}
\newcommand{\ea}{\end{eqnarray}}

\def\lsim{~\rlap{$<$}{\lower 1.0ex\hbox{$\sim$}}}

\def\gsim{~\rlap{$>$}{\lower 1.0ex\hbox{$\sim$}}}

\title[The HI mass function as a probe of photoionisation feedback on low mass galaxy formation]
      {The HI mass function as a probe of photoionisation feedback on low mass galaxy formation}
      \author[Han-Seek Kim et al.]
      {Han-Seek~Kim$^{1}$\thanks{hansikk@unimelb.edu.au}, J. Stuart. B.~Wyithe$^{1,3}$, C.~Power$^{2,3}$,  Jaehong~Park$^{1}$, C. d. P.~Lagos$^{2}$, \vspace{0.3cm} \\ 
      \LARGE\rm{and C. M.~Baugh$^{4}$} \\
       $^1$School of Physics, The University of Melbourne, Parkville, VIC 3010, Australia\\
       $^2$International Centre for Radio Astronomy Research, 
                 University of Western Australia, 35 Stirling Highway, 
                 Crawley, WA 6009, Australia\\
       $^3$ARC Centre of Excellence for All-sky Astrophysics (CAASTRO)\\
      $^4$Institute for Computational Cosmology, Department of Physics, University of Durham, South Road, Durham DH1 3LE, UK}
  	
\date{}

\pagerange{\pageref{firstpage}--\pageref{lastpage}}
\pubyear{2014} 

\begin{document}

\maketitle

\label{firstpage}

\begin{abstract}
We explore the galaxy formation physics governing the low mass end of the HI mass function in the local Universe.
Specifically, we predict the effects on the HI mass function of varying i) the strength of photoionisation feedback and the redshift of the end of the epoch of reionization, ii) the cosmology, iii) the supernovae feedback prescription, and iv) the efficiency of star formation.
We find that the shape of  the low-mass end of the HI mass function is most affected by the critical halo mass below which galaxy formation is suppressed by photoionisation heating of the intergalactic medium. We model the redshift dependence of this critical dark matter halo mass by requiring a match to the low-mass end of the HI mass function. The best fitting critical dark matter halo mass decreases as redshift increases in this model, corresponding to a circular velocity of $\sim  50 \, {\rm km \,s}^{-1}$ at $z=0$, $\sim  30 \, {\rm km\, s}^{-1}$ at $z \sim 1$ and 
$\sim 12 \, {\rm km \, s}^{-1}$ at $z=6$. We find that an evolving critical halo mass is required to explain both the shape and abundance of galaxies in the HI mass function below $M_{\rm HI} \sim 10^{8} h^{-2} {\rm M_{\odot}}$. The model makes specific predictions for the clustering strength of HI-selected galaxies with HI masses $> 10^{6} h^{-2} {\rm M_{\odot}}$ and $> 10^{7} h^{-2} {\rm M_{\odot}}$ and for the relation between the HI and stellar mass contents of galaxies which will be testable with upcoming surveys with the Square Kilometre Array and its pathfinders. We conclude that measurements of the HI mass function at $z \ge 0$ will lead to an improvement in our understanding of the net effect of photoionisation feedback on galaxy formation and evolution.
\end{abstract}


\section{Introduction}

There is a wealth of observational evidence that the Universe experienced an ``Epoch of
Reionization'' (EoR) at $z \geq 6$
\citep[e.g.][]{ouchi.etal.2010,mesinger.2010,mcgreer.etal.2011,shull.etal.2012}
during which time the cosmic  diffuse neutral hydrogen was ``re-ionized'' by a 
background of ultra-violet (UV) and X-ray 
radiation produced by the first generation of stars and galaxies 
\citep[e.g.][]{barkana.loeb.2007,robertson.etal.2010}.
This photo-ionizing background heated the gas in low-mass dark matter halos, leading to
its expulsion from their shallow potential wells \citep{Shapiro2004}. In addition, an 
increased Jeans mass prevented infall of gas onto low mass halos with virial temperatures $\leq 10^4$ K \citep[cf.][]{efstathiou.1992}. This implies that there should be a critical dark matter halo mass -- and therefore a halo
circular velocity -- below which galaxy formation should be suppressed during reionization. {Recently, \cite{Brown2014} presented high-resolution spectroscopy of stars in six ultra-faint dwarfs within the vicinity of the Milky-Way and analysed their stellar population properties. Brown et~al. concluded that these galaxies formed $\sim$ 80\% of their stars by $z \sim$ 6, and therefore are the best nearby candidates for probing star formation quenching during the EoR. These galaxies have maximum velocities $\sim 10 \, {\rm km\, s}^{-1}$, imposing an observational limit for star formation quenching due to the reionization.}

\citet{rees.1986} used analytical arguments to deduce that this critical circular velocity should be
$V_{\rm cut} \simeq 30 \, {\rm km\, s}^{-1}$, reasoning that the rate at which gas is heated by an ionizing 
background is balanced by the rate at which it cools radiatively. Subsequent studies have tried
to sharpen the prediction for $V_{\rm cut}$, with calculations based on semi-analytic approaches 
\citep[e.g.][]{efstathiou.1992,benson.etal.2002}, idealised 1D hydrodynamical calculations 
\cite[e.g.][]{thoul.weinberg.1996} and fully cosmological hydrodynamical simulations 
\citep[e.g.][]{gnedin.2000,okamoto.etal.2008}. However, there is still debate about the precise
value of $V_{\rm cut}$. Recently, \cite{sobacchi.etal.2013} used 1D collapse simulations to show that the critical halo mass depends on the redshift,
following earlier work by \cite{dijkstra2004}.  

The HI mass function in the local Universe provides information about galaxy formation physics in low mass dark matter halos. Indeed, 
\cite{kim2013a} recently showed that feedback effects are important in determining the HI mass function in the local Universe.
In particular, the low mass end of the HI mass function is governed by photoionisation feedback processes in small dark matter halos. 
In this paper we discuss the question of whether such modelling can be used to gain further insight into the value of $V_{\rm cut}$. Specifically we explore the effect of photoionisation feedback on the HI mass function using the GALFORM semi-analytic galaxy formation model \cite[]{cole.etal.2000}. This 
model reproduces the observed HI mass function at $z$=0, accurately 
matching its amplitude and shape at intermediate and high HI masses.
We compare the models with observational estimates of the HI mass function
from the HIPASS \citep[HI Parkes All-Sky Survey; see][]{meyer.etal.2004} 
and ALFALFA \citep[Arecibo Legacy Fast ALFA Survey; see][]{giovanelli.etal.2005}
surveys. We use the version of the semi-analytical galaxy formation model 
{GALFORM} 
developed by
 \citet{Lagos2011,lagos.etal.2011b} as the default model. 
 
The structure of the paper is as follows. In \S\ref{sec:model}, we provide
a brief overview of the \citet{Lagos2011} model.
In \S\ref{sec:himf} we show how the different ingredients influence 
the HI mass function. In \S\ref{sec:modified} we describe the modified model for photoionisation feedback. We finish with a summary of our results in \S\ref{sec:summary} 

\section{The Galaxy Formation Model} 
\label{sec:model}

We use the semi-analytical galaxy formation model {GALFORM} 
\citep[cf.][]{cole.etal.2000,baugh.2006} described in \citet{Lagos2011,lagos.etal.2011b} 
to predict the properties of galaxies forming and evolving in the
$\Lambda$CDM cosmology\footnote{Recall that the cosmological
parameters adopted for the Millennium Simulation are the total matter density
$\Omega_{\rm M}=0.25$, the baryon density $\Omega_{\rm b}=0.045$, 
the vacuum energy density $\Omega_{\Lambda}=0.75$, the Hubble parameter
$H_{0}=73 \,{\rm km \, s^{-1}}\,{\rm Mpc}^{-1}$, the primordial scalar spectral 
index $n_{\rm s}=1$ and the fluctuation amplitude $\sigma_{8}=0.9$.}. 
GALFORM models the key physical 
processes of galaxy formation, including the gravitationally driven assembly 
of dark matter halos, radiative cooling of gas and its collapse to form 
centrifugally supported discs, star formation, and feedback from
supernovae (SNe) and active galactic nuclei (AGN).

\citet{Lagos2011} extended GALFORM by modelling the splitting
of cold gas in the interstellar medium (ISM) into its HI and H$_2$ components 
and by explicitly linking star formation to the amount of H$_2$ present in a 
galaxy. \citet{Lagos2011} compared empirically and theoretically 
derived star formation laws \citep[cf.][]{blitz.2006,krumholz.etal.2009} with 
a variety of observations (e.g. the HI mass function, $^{12}$CO (1-0) 
luminosity function, and correlations 
between the ratio H$_2$/HI and stellar and cold gas masses in \cite{lagos.etal.2011b}) and found that the
empirical law of \citet{blitz.2006} (see also \citealt{leroy.etal.2008}) is 
favoured by these data. This law is of the form
\begin{equation}
\Sigma_{\rm SFR} = \nu_{\rm SF} \,\rm f_{\rm mol} \, \Sigma_{\rm gas},
\label{Eq.SFR}
\end{equation}
\noindent where $\Sigma_{\rm SFR}$ and $\Sigma_{\rm gas}$ are the surface
densities of the star formation rate (SFR) and total cold gas mass 
respectively, $\nu_{\rm SF}$ is the inverse of the SF timescale for the 
molecular gas and $\rm f_{\rm mol}=\Sigma_{\rm mol}/\Sigma_{\rm gas}=R_{\rm mol}/(R_{\rm mol}+1)$ is the
molecular to total gas mass surface density ratio. 
$R_{\rm mol}$ is defined by $R_{\rm mol}=\Sigma(H_{2})/\Sigma(HI)=(P_{\rm ext}/P_{\rm 0})^{\beta_{\rm press}}$. 
{$P_{\rm ext}$ is the interstellar gas pressure in a galaxy. We calculate $P_{\rm ext}$ using the approximation from \cite{Elmegreen1993},
\begin{equation}
P_{\rm ext}\approx{\pi \over 2}G\,\Sigma_{\rm gas}\left[\Sigma_{\rm gas}+\left({\sigma_{g} \over \sigma_{*}}\right)\Sigma_{*}\right],
\end{equation}
where $G$ is the gravitational constant and $\Sigma_{*}$ is the total surface densities of the stars.
The velocities dispersions of the gas and stars are given, respectively, by $\sigma_{\rm gas}$ 
and $\sigma_{\rm *}$.} 
The values allowed by the observations of \cite{{leroy.etal.2008}} and \cite{Bi2010} are $\nu_{\rm SF}=0.52\pm0.25~{\rm Gyr}^{-1}$, $\rm {log}(P_{\rm 0}/k_{\rm B}[cm^{-3}K])~=~4.54\pm0.07$ 
and $\beta_{\rm press}=0.92\pm0.07$.
Importantly for the work
we present in this paper, \citet{Lagos2011HI} showed that the 
\citet{blitz.2006} law is able to broadly reproduce the HI mass function at $z$=0 at 
high and intermediate HI masses, $M_{\rm HI}>10^{8}h^{-2} {\rm M_{\odot}}$. This is because it suppresses star formation 
in lower mass galaxies due to their low $\Sigma_{\rm gas}$, thereby reducing SNe feedback and allowing these 
galaxies to retain larger gas reservoirs. Note that we use the 
\citet{lagos.etal.2011b} model as the default model in this paper. In addition to the \citet{blitz.2006} star formation law, this model has longer duration 
starbursts compared with the model of \cite{Bower2006}.

As discussed in \citet{kim2013a}, photo-ionization is predicted to have a dramatic 
impact on star formation in low-mass galaxies. This is because the presence of a 
photo-ionizing background 
both modifies the net cooling rate of gas in halos by removing the ``hydrogen peak''
in the cooling curve \citep[cf. Fig. 1 of][]{benson.etal.2002} and increases
the temperature of the intergalactic medium (IGM) such that its thermal pressure 
prevents gravitational collapse onto low-mass halos 
\citep[e.g.][]{efstathiou.1992,okamoto.etal.2008}. As a result, only those halos that 
hosted cold gas prior to re-ionization can form stars 
\citep[e.g.][]{hoeft.etal.2006}. {{GALFORM}} includes the 
\citet{benson.etal.2002} prescription for suppressing the cooling of 
halo gas onto the galaxy -- this occurs if the host halo's circular 
velocity $V_{\rm circ}$ lies below a threshold $V_{\rm cut}$ at redshift $z_{\rm cut}$.
The values in the default \citet{lagos.etal.2011b} model are 
$V_{\rm cut} = 30 \, {\rm km  \, s}^{-1}$ and $z_{\rm cut} = 10$. The default $V_{\rm cut}$ is motivated by the results of hydrodynamical simulations by \citet{hoeft.etal.2006} and 
\citet{okamoto.etal.2008}. However, several studies indicate that $V_{\rm cut}$ should be redshift dependent \citep{dijkstra2004,sobacchi.etal.2013}. The implementation of a single value of $z_{\rm cut}$ and $V_{\rm cut}$ may therefore be overly simplistic. We revise this modelling in this paper.

\section{Understanding the the low mass end of HI mass function in the local Universe}
\label{sec:himf}

The self-consistent treatment of HI in \citet{Lagos2011,lagos.etal.2011b} results in a model that shows good agreement between model predictions and the HIPASS and ALFALFA observations for HI masses greater than $M_{\rm HI}\sim10^{8} h^{-2}{\rm M_{\odot}}$. However the predicted HI mass function between 
$10^{6} h^{-2} {\rm M_{\odot}}$ and $10^{8} h^{-2} {\rm M_{\odot}}$ (i.e., the low mass end of the HI mass function) fails to capture the shape and abundance of the observed HI mass function. In this section we investigate how predictions of the low mass end of the HI mass function at $z\simeq$0 are influenced by the assumed photo-ionizing background feedback, cosmology, supernovae feedback prescription, and star formation law efficiency. We also introduce a description of the possible redshift dependence of the photo-ionizing feedback strength. Throughout the paper we use Monte-Carlo generated merger histories \citep{Parkinson2008}. We adopt a minimum halo mass of
$M_{\rm halo, min} = 5 \times 10^{8} \,h^{-1}\, {\rm M_{\odot}}$. From the chosen minimum halo mass, we can predict the neutral hydrogen gas mass structures down to the current observational limits \citep{martin.etal.2010}. 

\subsection{Photoionisation feedback at the end of reionization}

\begin{figure}
 \includegraphics[width=8.3cm]{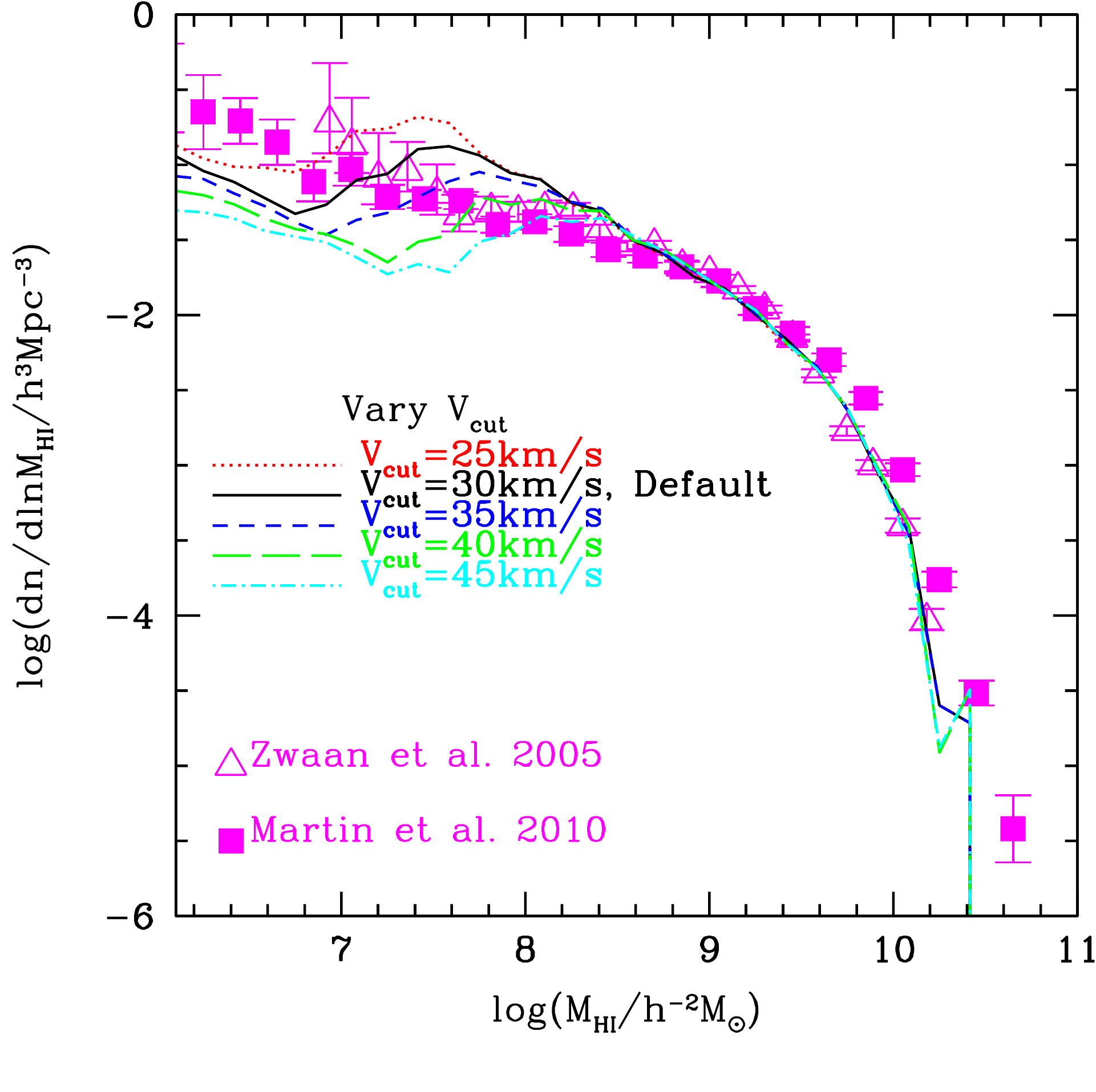}
    \includegraphics[width=8.3cm]{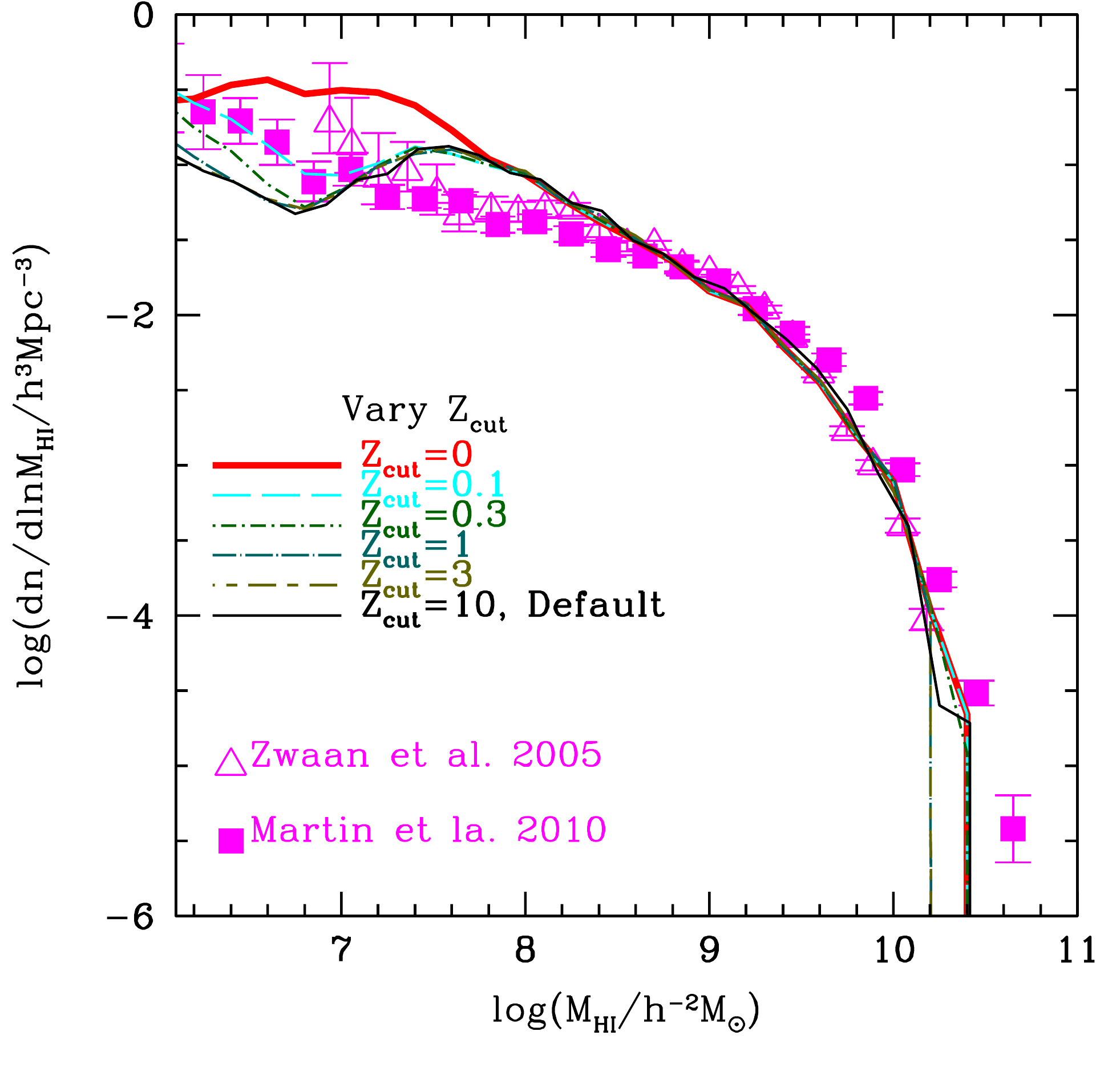}
  \caption{The impact of the photo-ionizing background on the predicted HI mass function. The panels show models for varying $V_{\rm cut}$ (top panel) and $z_{\rm cut}$ values (bottom panel). The
    symbols correspond to observations from
    HIPASS \citep[open triangles; cf.][]{Zwaan05} and ALFALFA
    \citep[filled squares; cf.][]{martin.etal.2010}. We vary the 
    photo-ionization strength between 
    $25 \, {\rm km \, s}^{-1}  \leq V_{\rm cut} \leq 45 \, {\rm km \, s}^{-1}$ (at a fixed $z_{\rm cut}$=10) and $0 \leq z_{\rm cut} \leq 10$ (at a fixed $V_{\rm cut}=30 \, {\rm km \, s}^{-1}$), as labelled in each panel.}
  \label{HIMFDE}
\end{figure}

In the standard implementation of {{GALFORM}}, the effect of photoionisation feedback induced by the epoch of reionization is modelled
by imposing a circular velocity cut-off $V_{\rm cut}$  on galaxy formation at a redshift of the end of reionization $z_{\rm cut}$.
In Fig.~\ref{HIMFDE} we show the resulting HI mass function assuming this standard model for reionization for a range of photoionisation strengths $V_{\rm cut}$ (at fixed $z_{\rm cut}$) and redshifts marking the end of reionizaiton $z_{\rm cut}$ (at fixed $V_{\rm cut} = 30 \, {\rm km \, s}^{-1}$).  These models can be compared with observations from ALFALFA \citep{martin.etal.2010} and HIPASS \citep{Zwaan05}. The single values of $V_{\rm cut}$ and $z_{\rm cut}$ introduce a non-monotonic feature in the HI mass function that is not present in the observations. The top panel of Fig.~\ref{HIMFDE} shows that the predicted abundance of galaxies with 2$\times$10$^{7}h^{-2} \, {\rm M_{\odot}} < M_{\rm HI} < 10^{8} h^{-2} {\rm M_{\odot} }$ is larger than observed when we choose the $V_{\rm cut} = 25 \, {\rm km \,s}^{-1}$, which matches the abundance at $M_{\rm HI} < 10^{7}\, h^{-2}\, {\rm M_{\odot}}$ (red dotted line). In contrast, the predicted abundance of galaxies with $M_{\rm HI} < 10^{7}\,h^{-2}\, {\rm M_{\odot}}$ is lower than observed when we select $V_{\rm cut} = 35\, {\rm km \, s}^{-1}$, which fits the abundance of galaxies with  $2 \times 10 ^{7}h^{-2} {\rm M_{\odot}} < M_{\rm HI} < 10^{8}\,h^{-2}\, {\rm M_{\odot}}$. This is evidence of the simple modelling of photoionisation feedback leading to the non-monotonic feature discussed above.

We also investigate the effect of varying the redshift of reionization at a fixed $V_{\rm cut} =\, {\rm 30 \,km \,s}^{-1}$. The bottom panel of Fig.~\ref{HIMFDE} shows that the value of $z_{\rm cut}$ does not influence the low mass end of the HI mass function for values ranging between $z_{\rm cut}$=1 to 10 (in the case of $V_{\rm cut} = \,30 \, {\rm km \,s}^{-1}$) indicating that the HI mass function in the local Universe cannot be used as a probe of the end of the epoch of reionization, at least using this simple modelling of photoionisation feedback.

\subsection{Other possibilities: cosmology, supernova feedback recipe, and star formation law}

\begin{figure}
    \includegraphics[width=7cm]{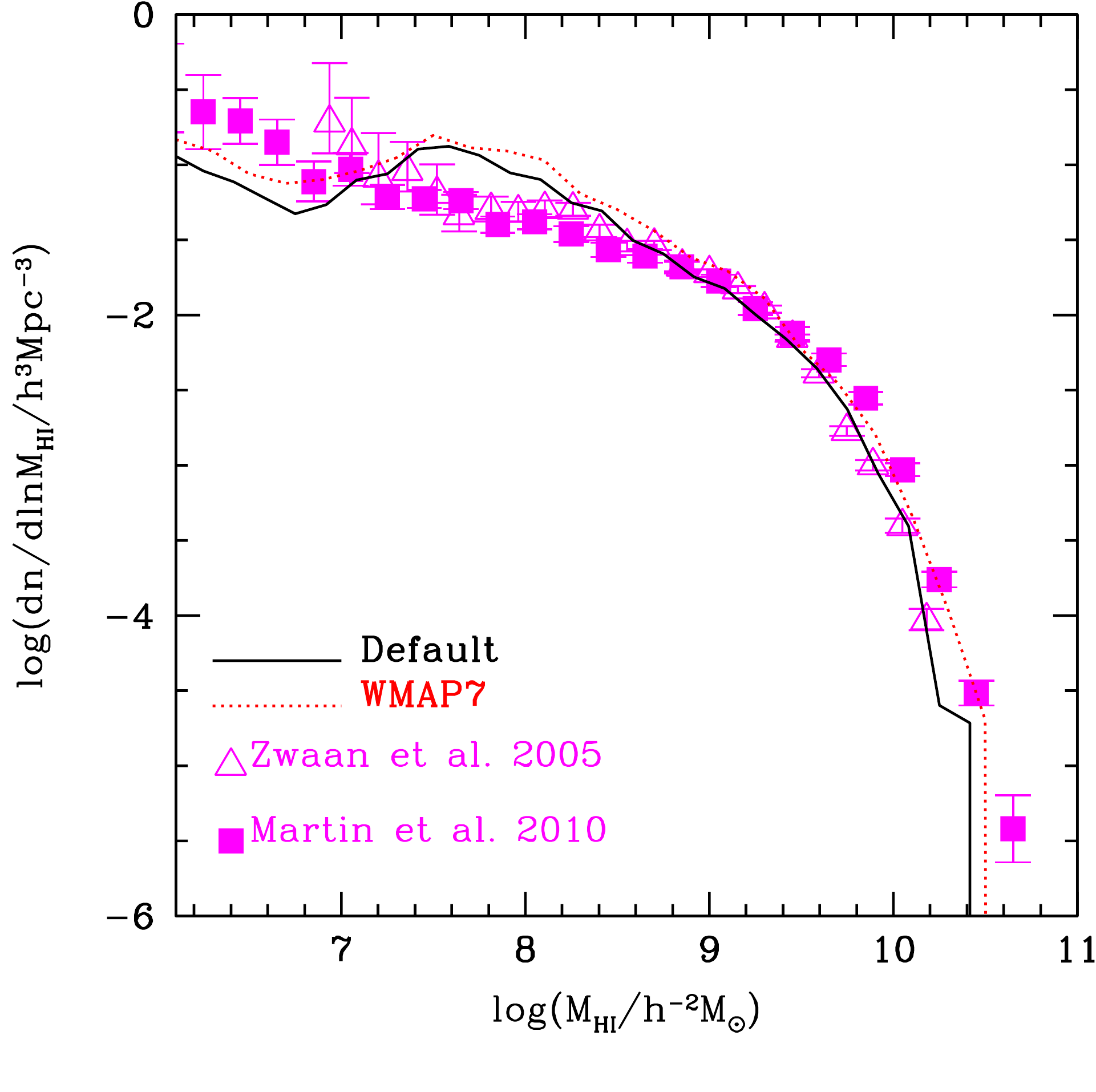}
    \includegraphics[width=7cm]{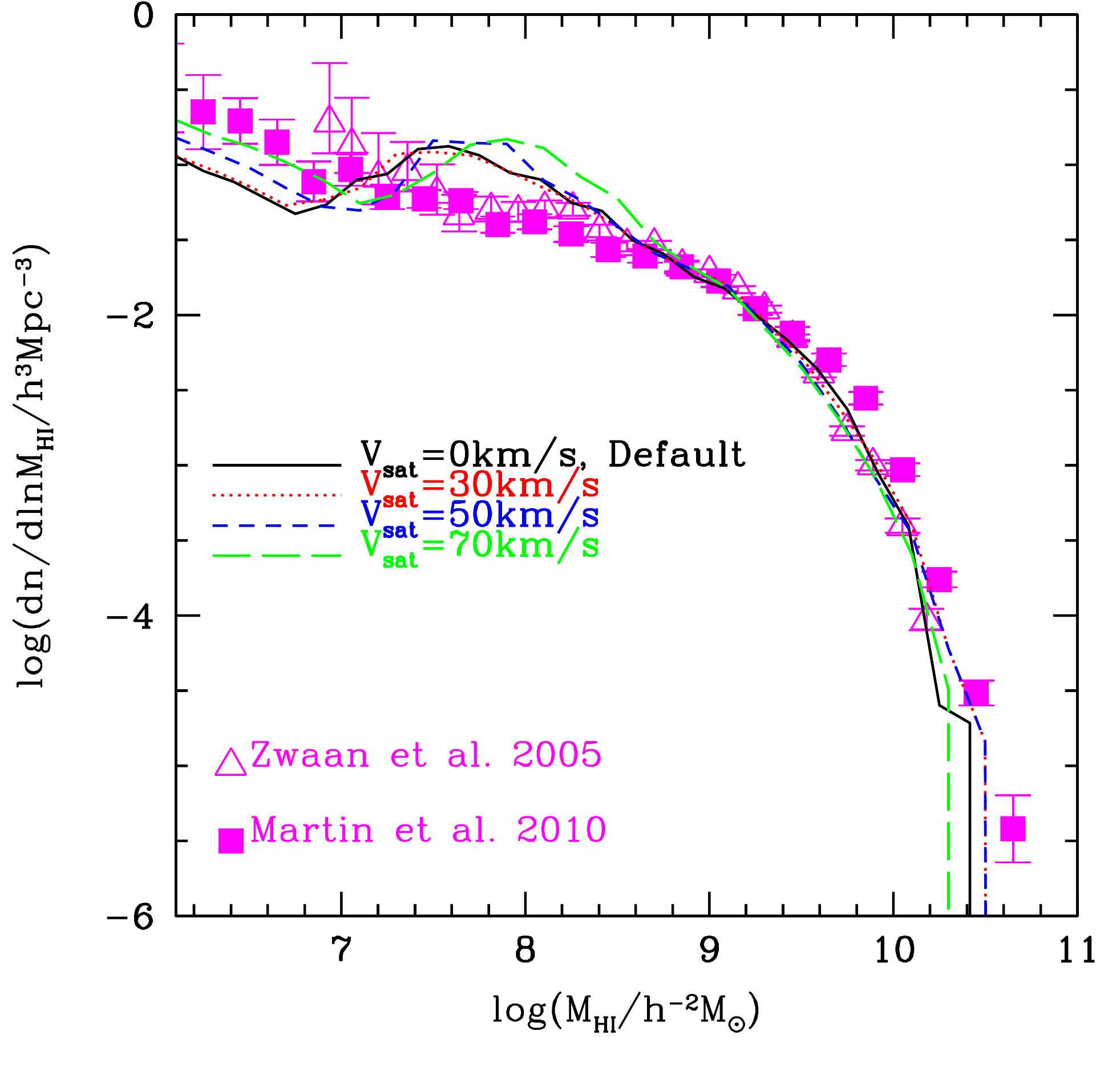}
    \includegraphics[width=7cm]{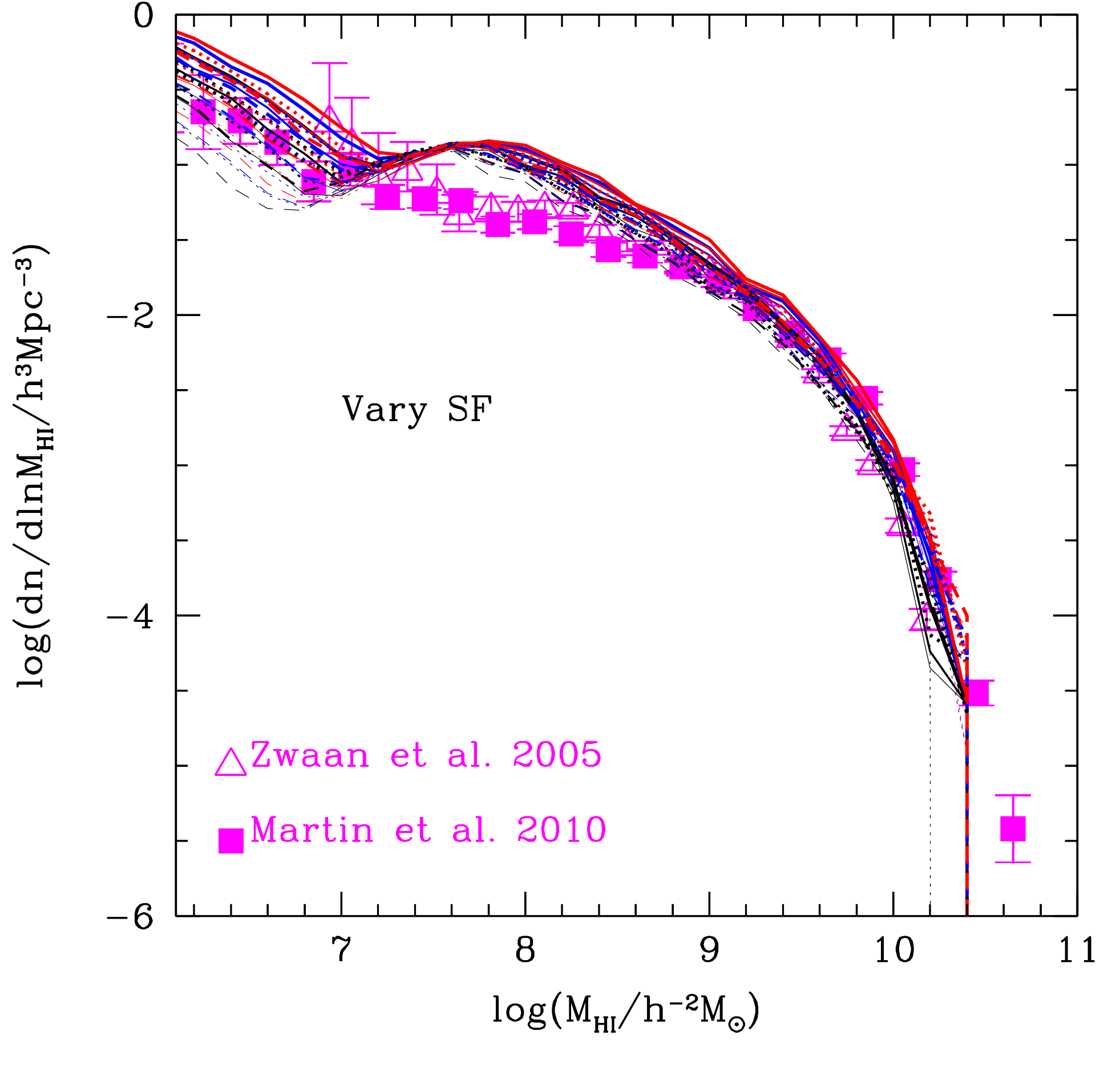}
  \caption{Impact of various galaxy formation physics on the HI mass function; Top panel: The effect of the changing the cosmology from WMAP1 to WMAP7. Middle panel: The effect of varying the circular velocity below which supernova feedback saturates, as described in \citet{Font.etal.2011}. Bottom panel: The effect of varying the parameters in the star formation law from \citet{lagos.etal.2011b} within the observational constraints (see test for details).  
    The symbols correspond to data from 
    HIPASS \citep[open triangles; cf.][]{Zwaan05} and ALFALFA
    \citep[filled squares; cf.][]{martin.etal.2010}.}
  \label{HIMFOthers}
\end{figure}
In this subsection we explore the effects of a range of additional aspects of galaxy formation on the HI mass function.

First, we inspect the impact of cosmology (top panel of Fig.~\ref{HIMFOthers}).
The default model has the cosmology used in the Millennium N-body simulation of \cite{Springel2005} as noted in Sec~\ref{sec:model}. This is compared with a model having the WMAP7 cosmology for which the parameters in GALFORM have been adjusted to reproduce a suite of other observations \citep{Violeta}. The abundance of galaxies across the full mass range of HI masses in the WMAP7 cosmology is larger than in the Millennium cosmology. 
However, the non-monotonic feature described in the last subsection remains and therefore is not dependent on cosmology. 

Second, we have explored the SNe saturation model suggested by \cite{Font.etal.2011}. The default GALFORM model parameterizes the SNe feedback mass loading efficiency as $\beta=(V_{\rm circ}/V_{\rm hot})^{-\alpha_{\rm hot}}$, where 
$V_{\rm circ}$ is the circular velocity of the galaxy at the half-mass radius. The parameters $V_{\rm hot}$ and $\alpha_{\rm hot}$ are adjustable 
and control the strength of SNe feedback. The default model has $V_{\rm hot} = 485 \, {\rm km\,s}^{-1}$ and $\alpha_{\rm hot}$=3.2 \cite[cf.][]{Bower2006}. In order to fit the number of satellite galaxies and their metallicity in the Milky Way, \cite{Font.etal.2011} proposed a scenario in which the SNe feedback efficiency is not a 
simple power law for all galaxy masses (as it is in the default model).  
They modelled an efficiency of SNe feedback that saturates in low-mass galaxies with 
circular velocities less than $65 \, {\rm km \,s}^{-1}$, but which follows the default model for larger galaxies \cite[see][for a physical interpretation of this saturation]{Lagos2013}.
Here we define the circular velocity of saturated low-mass galaxies as $V_{\rm sat}$, and explore
the impact on the HI mass function of varying $V_{\rm sat}$ (from $0-70 \,{\rm km \, s}^{-1}$, middle panel of Fig.~\ref{HIMFOthers}).
The abundance of galaxies with $M_{\rm HI} < 10^{7}\,h^{-2}\, {\rm M_{\odot}}$ increases as the value of $V_{\rm sat}$ increases, because less gas is reheated and expelled from the disk to the halo when the galaxy has circular velocity $V_{\rm circ}$ $\le$ $V_{\rm sat}$. 
However, as we found for the case of changing cosmology, the non-monotonic feature is still present in the predicted HI mass function even after varying $V_{\rm sat}$.

In addition, we explore the effect of changing the parameters of the star formation law within the observed range as noted in Sec.~\ref{sec:model}. The bottom panel of Fig.~\ref{HIMFOthers} shows the HI mass function resulting from varying values of $\nu_{\rm SF}$=[0.27, 0.5, 0.755] (solid-, dotted-, dashed-lines) ${\rm Gyr}^{-1}$, $P_{\rm 0}/k_{\rm B}$=[17000, 30000, 40738] (black-, blue-, red-line) $ {\rm cm}^{-3}K $
and $\beta_{\rm press}$=[0.8, 0.9, 0.99] (the thinest to the thickest). We show the HI mass functions from the sets of 27 combinations of three parameters of the star formation law. The varied $\nu_{\rm SF}$ and $P_{\rm 0}$ values lead to an overall shift in the HI mass function compared to the default model. Varying the value of $\beta_{\rm press}$ also produces an overall shift in the mass function. In addition, varying $\beta_{\rm press}$ results in a large change in abundance at the low mass end of the HI mass function. This change is most likely due to a lower $\Sigma_{H2}/\Sigma_{HI}$ in the regime $P_{\rm ext}<P_{\rm 0}$ (HI dominated regions) which is obtained for large values of $\beta_{\rm press}\sim$0.99. 

However, as in the cases of varying cosmology and SNe feedback, changing the SF law still results in a non-monotonic HI mass function for $M_{\rm HI} \le 10^{8} \,h^{-2}\,{\rm M_{\odot}}$ in all 27 cases, in disagreement with observations. This suggests that new or modified physics is needed in the default galaxy formation model in order to improve the agreement with the observations at $M_{\rm HI} \le 10^{8} \,h^{-2}\,{\rm M_{\odot}}${.} 
  
\section{An improved model for photo-ionisation feedback}\label{sec:modified}

In the previous section we explored a range of possibilities for phenomena which may govern the low and intermediate mass range of the HI mass function. 
In all cases the predictions include a non-monotonic feature which is not present in the observations. The results of Fig.~\ref{HIMFDE} suggest that the non-monotonic feature in the low mass end of the HI mass function is an artifact of adopting a single value of $V_{\rm cut}$ in the parametrization of photo-ionisation feedback. We therefore model 
photoionisation feedback using a redshift dependent $V_{\rm cut}(z)$ which is motivated by the calculations of \cite{sobacchi.etal.2013} using 1D collapse simulations. 
Note that in our default model, we use $\beta_{\rm press}=0.99$ in the star-formation law.
 
\cite{sobacchi.etal.2013} assumed the following functional form for the critical mass, $M_{\rm crit}$, defined as the total halo mass where the baryon fraction is half of the global value: 
\begin{equation}
M_{\rm crit}=M_{0}J^{a}_{21}\left(\frac{1+z}{10}\right)^{b}\left[1-\left(\frac{1+z}{1+z_{\rm IN}}\right)^{c}\right]^{d},\label{SME}
\end{equation}
where $M_{0}, a, b, c$ and $d$ are parameters that Sobacchi \& Mesinger fitted to the results of their numerical simulations. The square bracketed term of Eqn.~\ref{SME} is related to the redshift evolution of the UVB exposure and redshift. The parameter z$_{\rm IN}$ corresponds to the redshift at which the halo was exposed to a UV background. \cite{sobacchi.etal.2013} explored a large parameter space of halo mass ($M_{0}$), UV background intensity ($J_{21}$, expressed in units of $10^{-21} {\rm erg \, s^{-1} \, Hz^{-1} cm^{-2} \, sr^{-1}}$), redshift ($z$), and redshift of UV background exposure ($z_{\rm IN}$) of galaxies. These are uncertain because the details of how reionization proceeded are yet to be observed. 

\cite{sobacchi.etal.2013} quoted the best fitting values of the parameters in Eqn.~\ref{SME} to their simulation results as ($M_{0}, a, b, c, d$)=($2.8 \times 10^{9}\, {\rm M_{\odot}}, 0.17, -2.1, 2.0, 2.5$). For our purposes, it is helpful to recast the critical halo mass as a circular velocity,  $V_{\rm cut}$,\citep{cole.etal.1994} 

\begin{equation}
M_{\rm crit}=3.36 \times 10^{5} \left(\frac{V_{\rm cut}}{{\rm km\, s^{-1}}}\right)^{3}\, (1+z)^{-\frac{3}{2}}\, {\rm M_{\odot}} ,\label{MHALO}
\end{equation} 
where we have assumed a Hubble constant of $70 \,{\rm km \,s}^{-1} {\rm Mpc}^{-1}$.
Combining Eqns.~(\ref{SME}) \&~(\ref{MHALO}), we obtain a redshift dependent $V_{\rm cut}(z)$, which, after adopting the suggested best fitting 
values for the parameters $b, c$ and $d$ for Eqn.~(\ref{SME}) is given by 
\begin{equation}
\begin{split}
V_{\rm cut}(z)  [{\rm km \, s}^{-1}] = V_{\rm cut0}(1+z)^{\alpha_{v}}
                   \left[1-\left(\frac{1+z}{1+z_{\rm IN}}\right)^{2}\right]^{2.5/3}.
\end{split}
\label{VZ}
\end{equation}
Here the $V_{\rm cut0}$ is the circular velocity of dark matter halos at z=0 below which photoionisation feedback suppresses gas cooling and 
$\alpha_{v} = b/3 + 1/2$ in the notation of Sobacchi \& Mesinger.  For the best fitting value of $b=-2.1$, $\alpha_{v} = -0.2$. 
For simplicity, we assume the redshift of UV background exposure to be z$_{\rm IN}$=10.  
The value of $V_{\rm cut0}$ is $ \approx 100 \, {\rm km \,s}^{-1}$ if we set $J_{21}=1$. If we use this value of $V_{\rm cut0}$, the model underpredicts 
the low-mass end of the HI mass function by a substantial amount and also underpredicts the faint-end of the galaxy luminosity function 
in the optical. We find that the model gives a better match to the observed HI mass function if we instead use a critical mass at $z=0$ 
which is reduced by a factor of 10 with respect to the value advocated by Sobacchi \& Mesinger, corresponding 
to $V_{\rm cut0} \approx 50 \,{\rm km \, s}^{-1}$.  This difference could arise due to the 1D simulations used by Sobacchi \& Mesinger  
missing 3D substructures that favour a lower value for the critical mass at z=0 \cite[see Sec 3.1 in][]{sobacchi.etal.2013}. 
In what follows, we treat $V_{\rm cut0}$ and $\alpha_{v}$ as free parameters in the new photo-ionisation feedback model. 

\begin{figure*}
  \includegraphics[width=7.5cm]{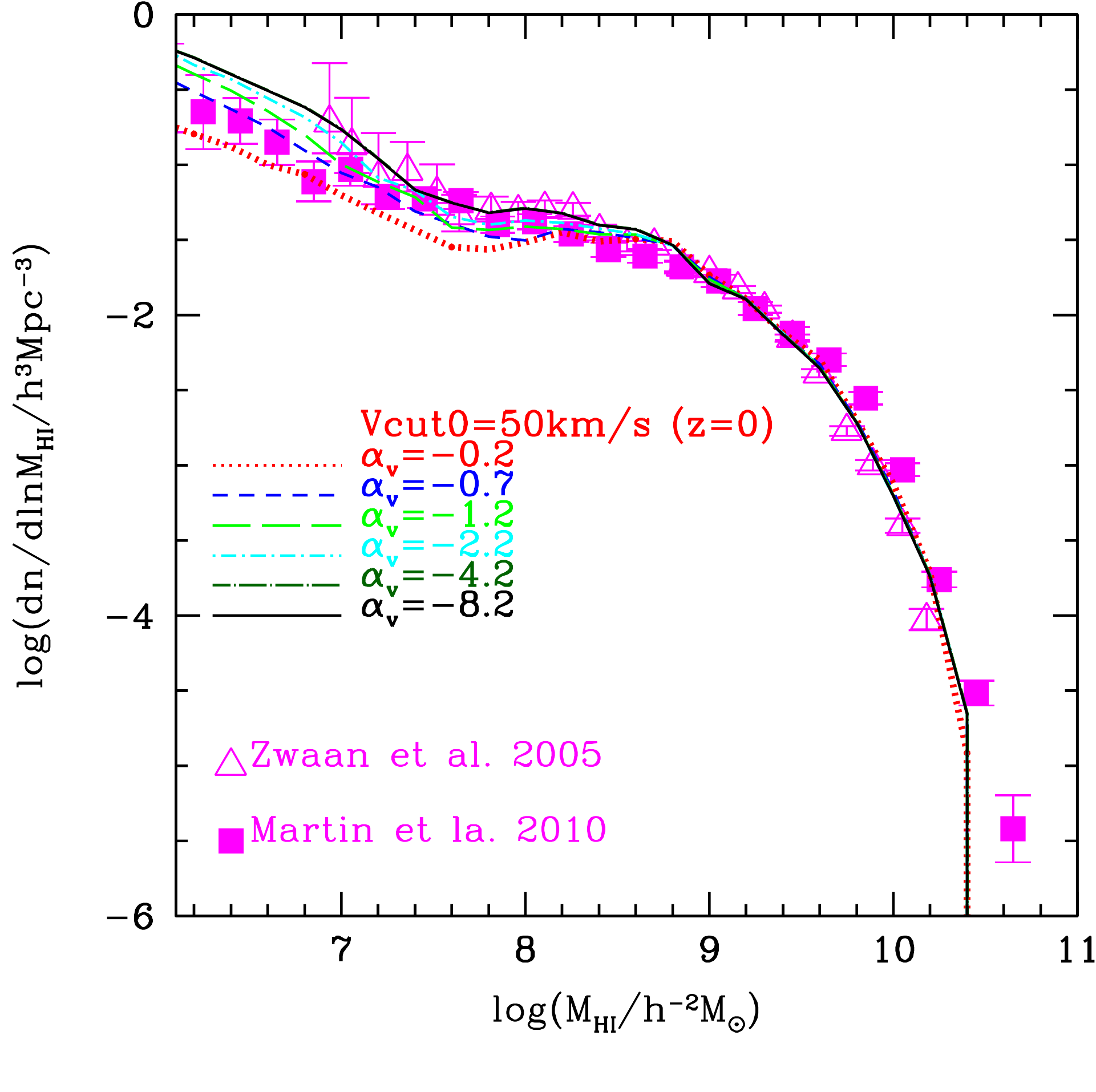}
   \includegraphics[width=7.5cm]{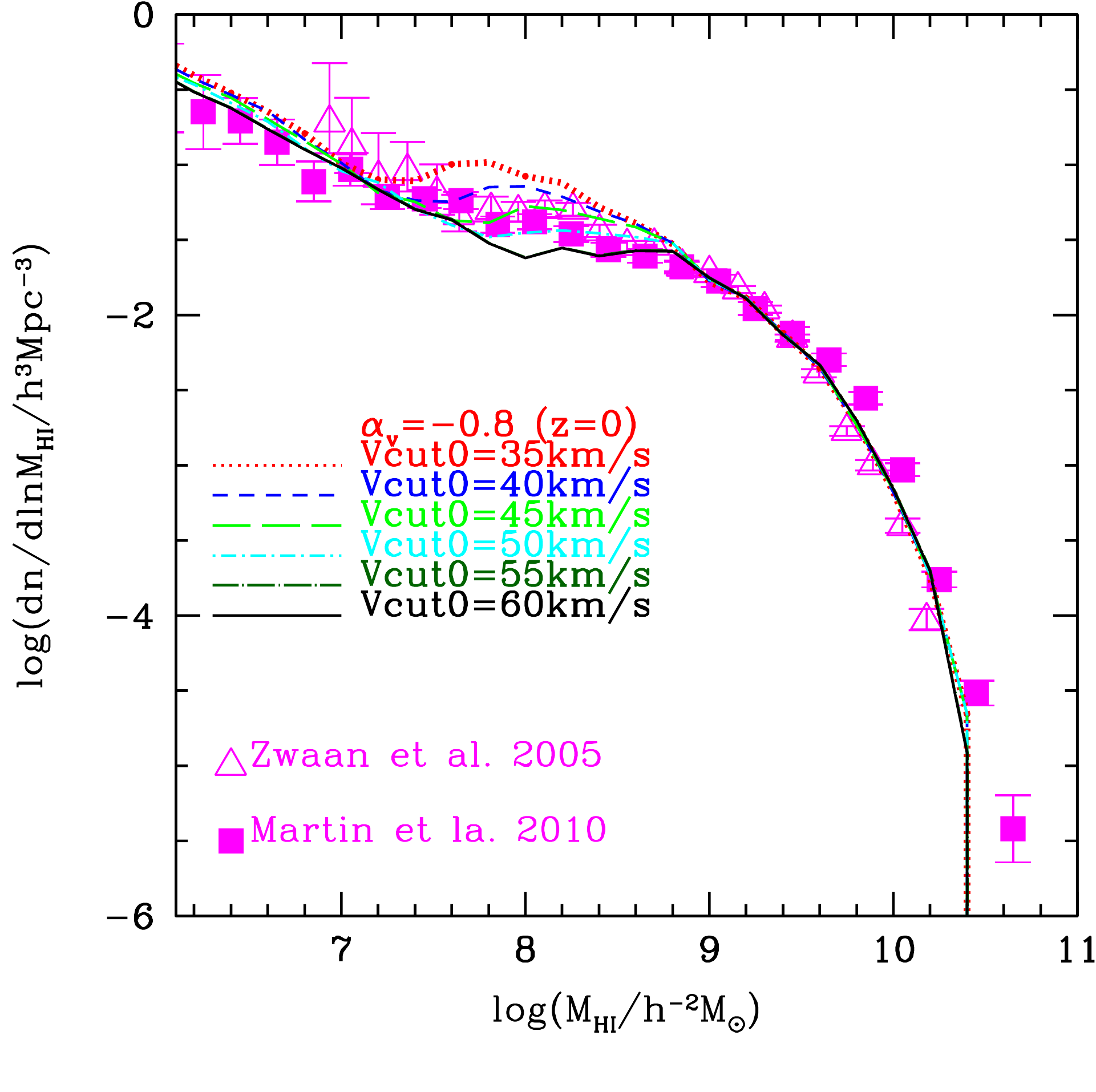}
   \includegraphics[width=7.5cm]{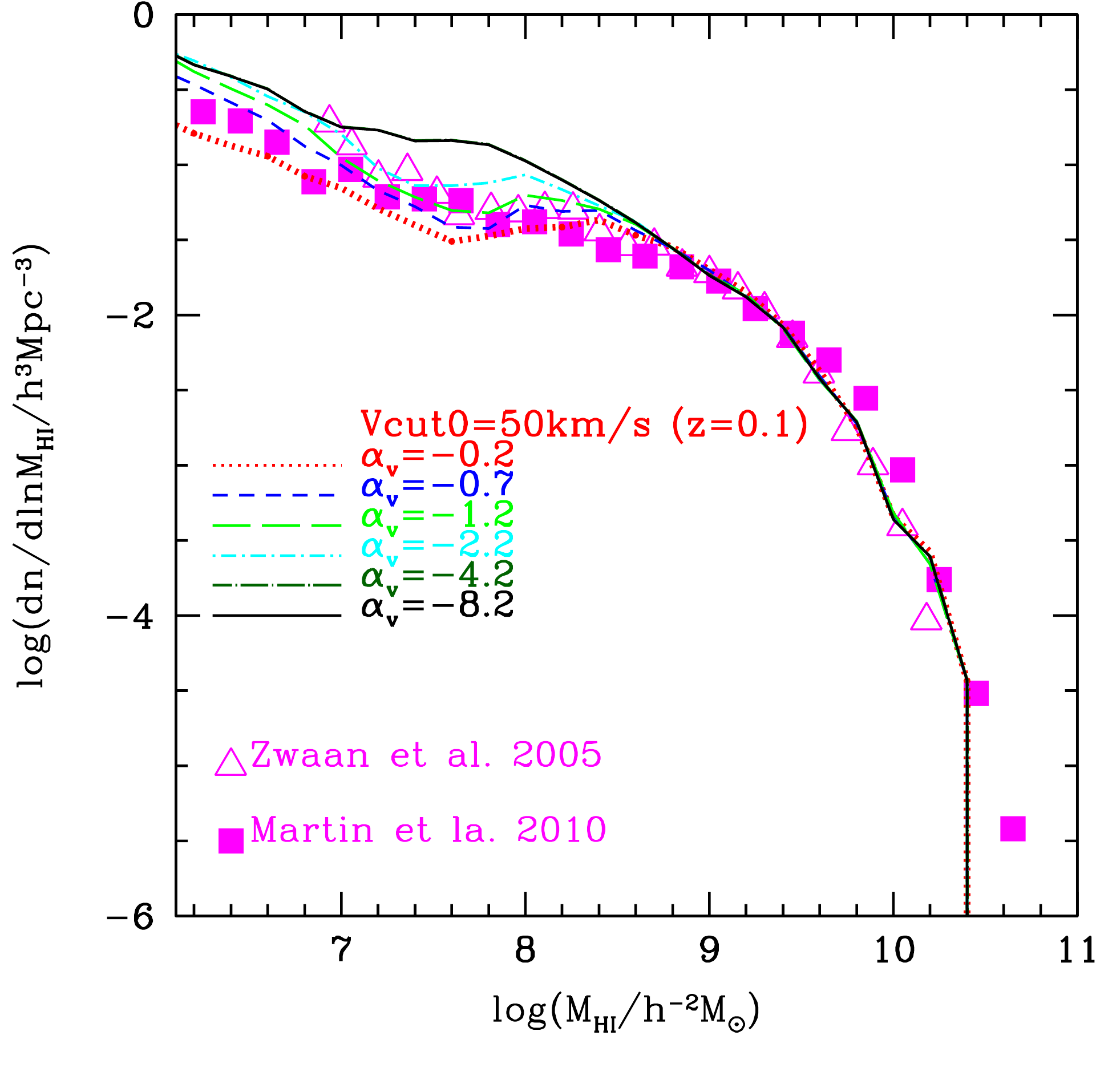}
\includegraphics[width=7.5cm]{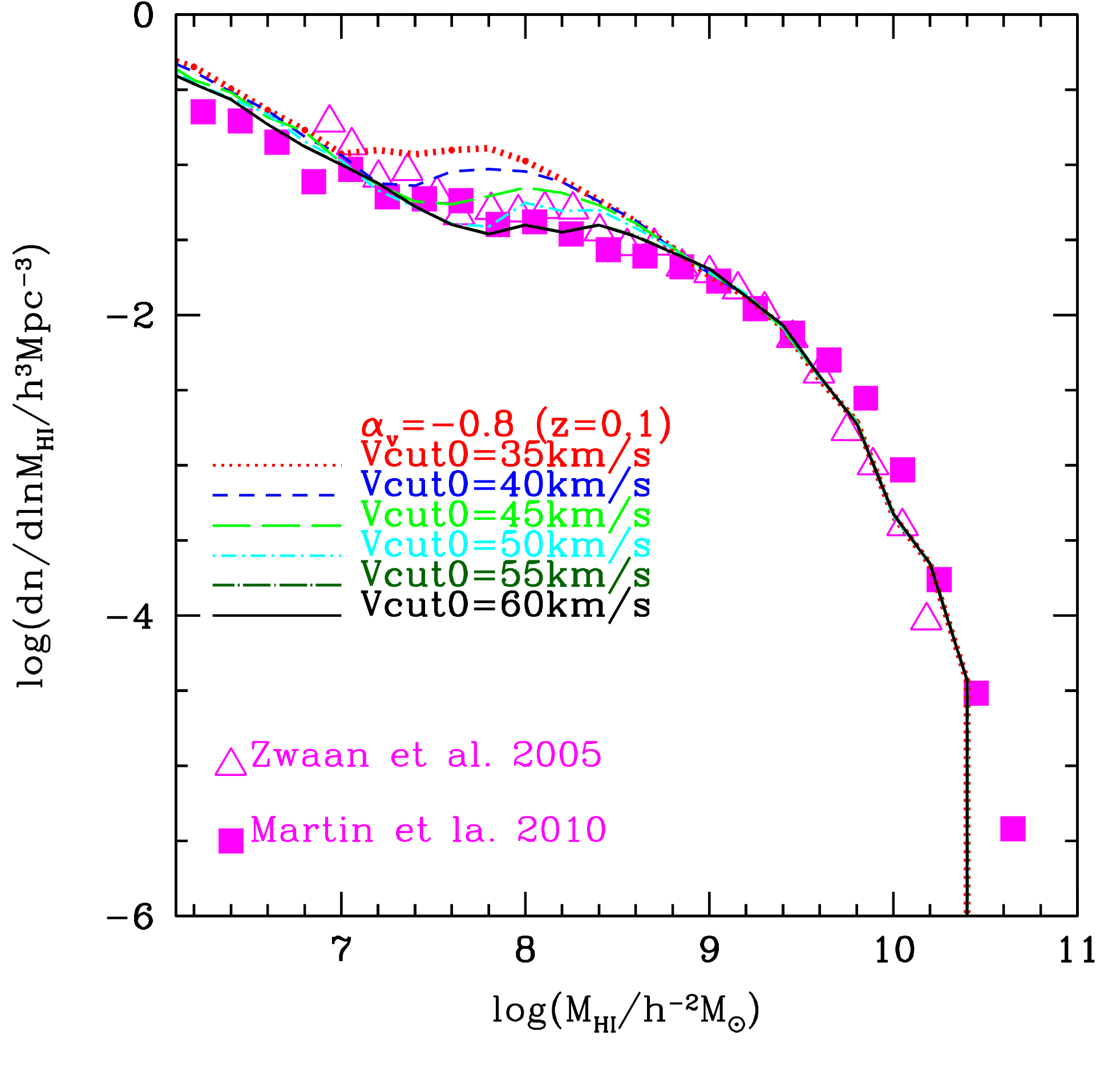}
\includegraphics[width=7.5cm]{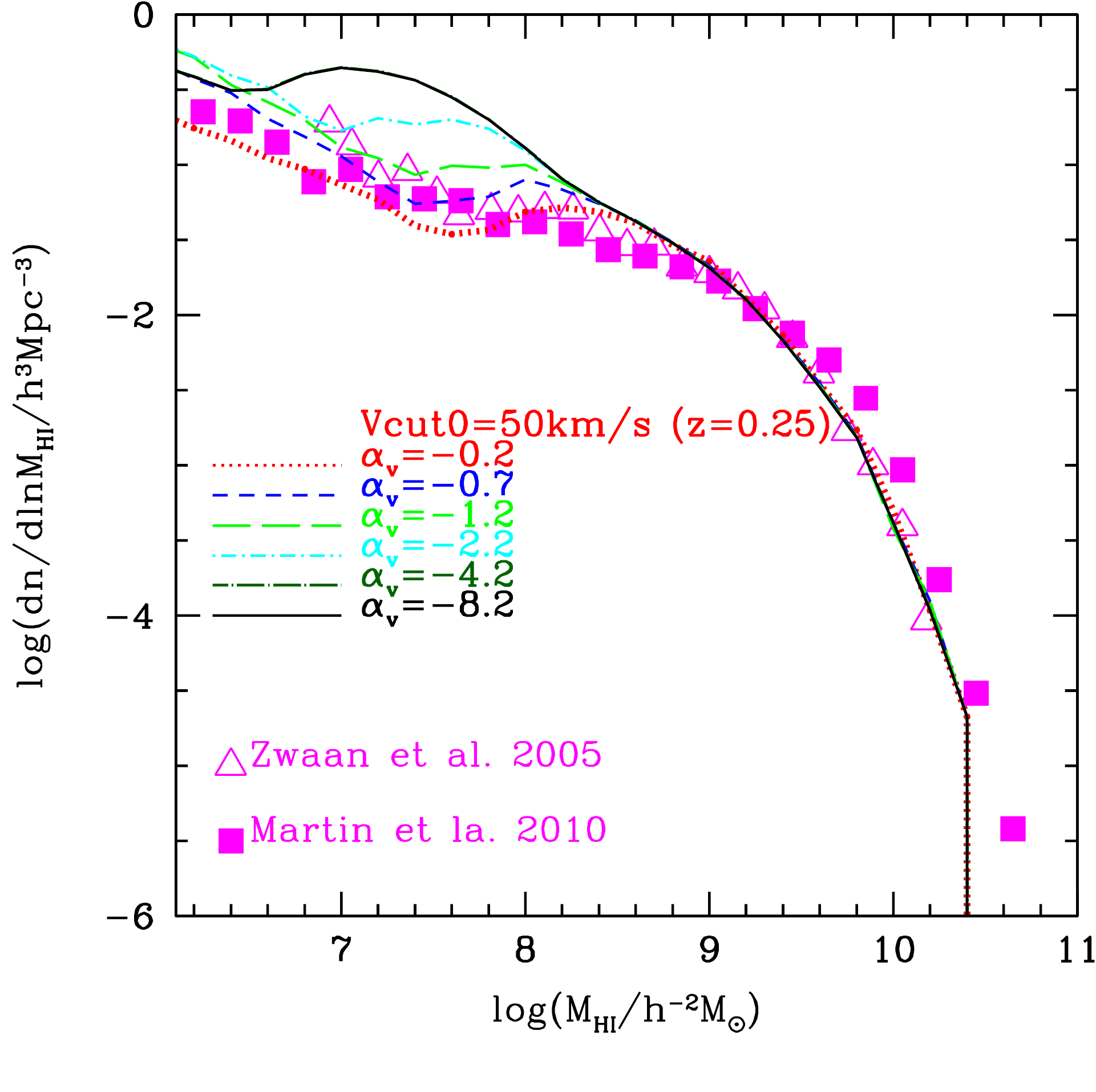}
\includegraphics[width=7.5cm]{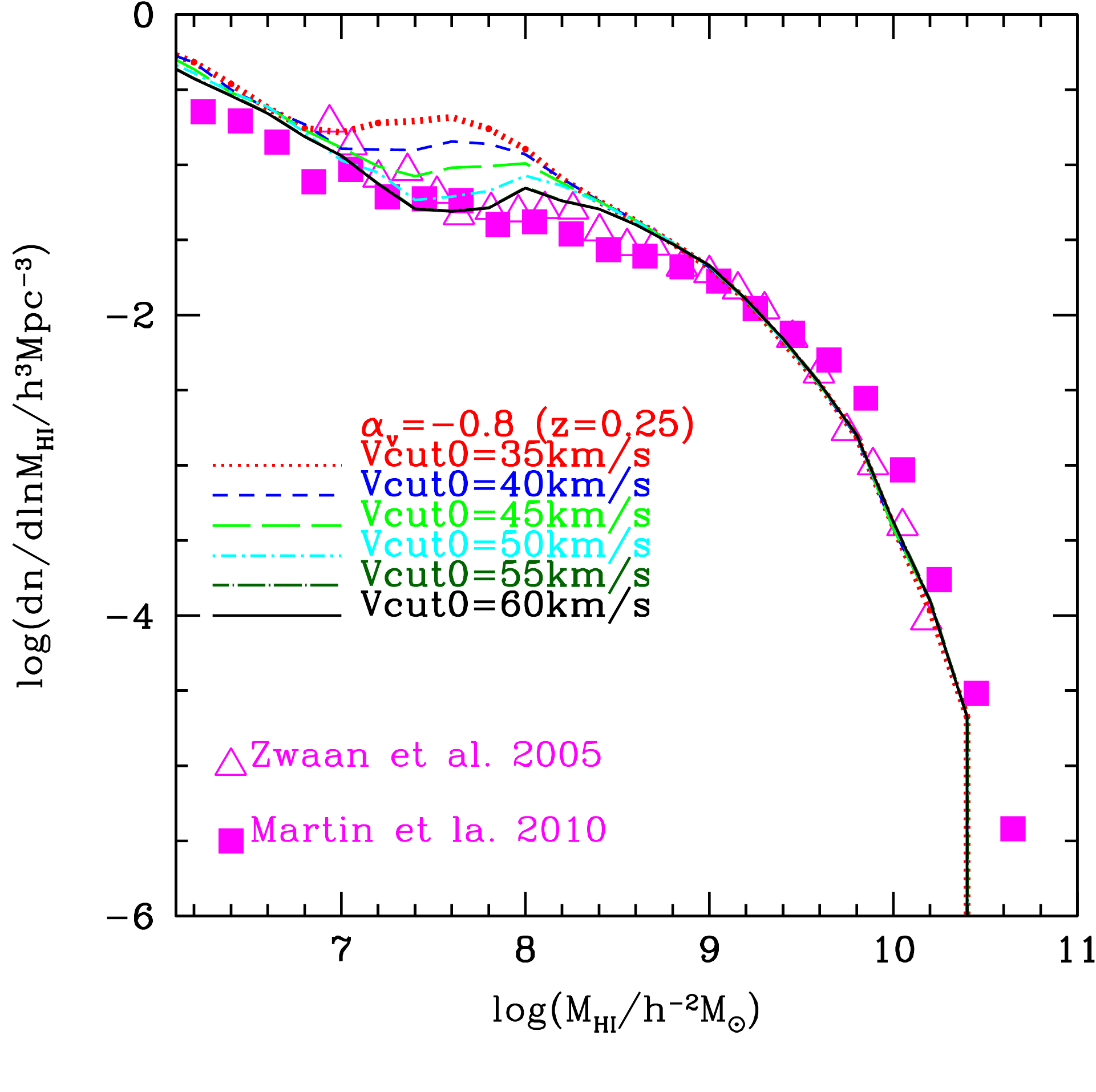}\vspace{-0.4cm}
  \caption{The impact of the redshift dependent $V_{\rm cut}(z)$ modelling on the predicted HI mass function (top panels) at z=0.
    The symbols correspond to observations in the local Universe from 
    HIPASS \citep[open triangles; cf.][]{Zwaan05} and ALFALFA
    \citep[filled squares; cf.][]{martin.etal.2010}. Note that these data are reproduced without error bars in the z=0.1 and z=0.25 middle and bottom panels as a reference to illustrate the evolution of the mass function. We vary $\alpha_{v}$ from -8.2 to -0.2 (which controls the redshift dependence), and $V_{\rm cut0}$ from $35 -60 \, {\rm km \,s}^{-1}$, as labelled. The middle and bottom panels show the predicted HI mass functions at z=0.1 and z=0.25, respectively, for the same combination of parameters shown in the top panels.}
  \label{HankPHZ}
\end{figure*}

In the top left panel of Fig.~\ref{HankPHZ} we show the predicted HI mass function at $z=0$ for models using $V_{\rm cut}(z)$ (Eqn.~\ref{VZ}) with $V_{\rm cut0}= 50 \, {\rm km \, s}^{-1}$, while varying $\alpha_{v}$ from -8.2 to -0.2. In the top right panel of Fig.~\ref{HankPHZ}, we show the predicted HI mass function for models using $V_{\rm cut}(z)$ and $\alpha_{v}$=-0.8, while varying $V_{\rm cut0}$ from $35 - 60 \, {\rm km \,s}^{-1}$ at z=0. The top panels of Fig.~\ref{HankPHZ} show that this model is able to capture the shape of the HIMF down to $M_{\rm HI}\sim \, 10^{6}\,h^{-2}\, {\rm M_{\odot}}$ much more accurately than the default model does at z=0. This suggests that an evolving photoionisation model is preferred by the observations in the local Universe.

We also show the predicted HI mass function at z=0.1 (middle panels) and z=0.25 (bottom panels) in Fig.~\ref{HankPHZ}. Interestingly, the predicted HI mass functions at these redshifts show larger variation on changing the parameters of  Eqn.~(\ref{VZ}) than the HI mass function at z=0. Ongoing and future HI-selected galaxy surveys using the SKA \cite[Square Kilometre Array; e.g.][]{Baugh2004SKA,Power2010,Kim2011} and its pathfinders, such as ASKAP \cite[Australian Square Kilometre Array Pathfinder; cf.][]{ASKAP2008}, and MeerKAT \cite[Meer Karoo Array Telescope; cf.][]{MeerKAT2007} are expected to extend our view of the HI Universe to higher redshifts. These observations will probe the physical processes that drive galaxy formation, and we expect these surveys to be particularly enlightening regarding how photoionisation feedback should be modelled.   
  
 \begin{figure*}
\includegraphics[width=8.4cm]{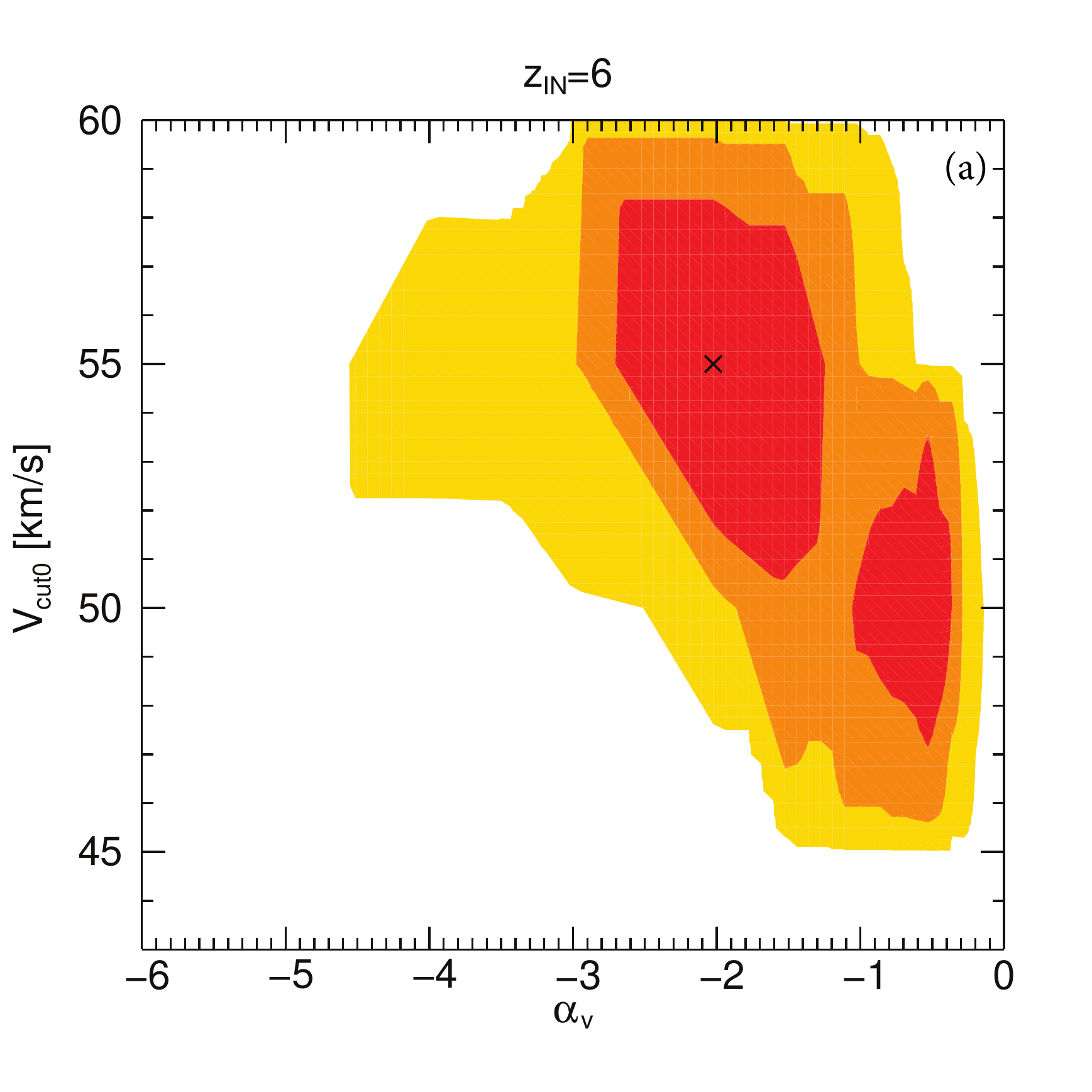}
\includegraphics[width=8.4cm]{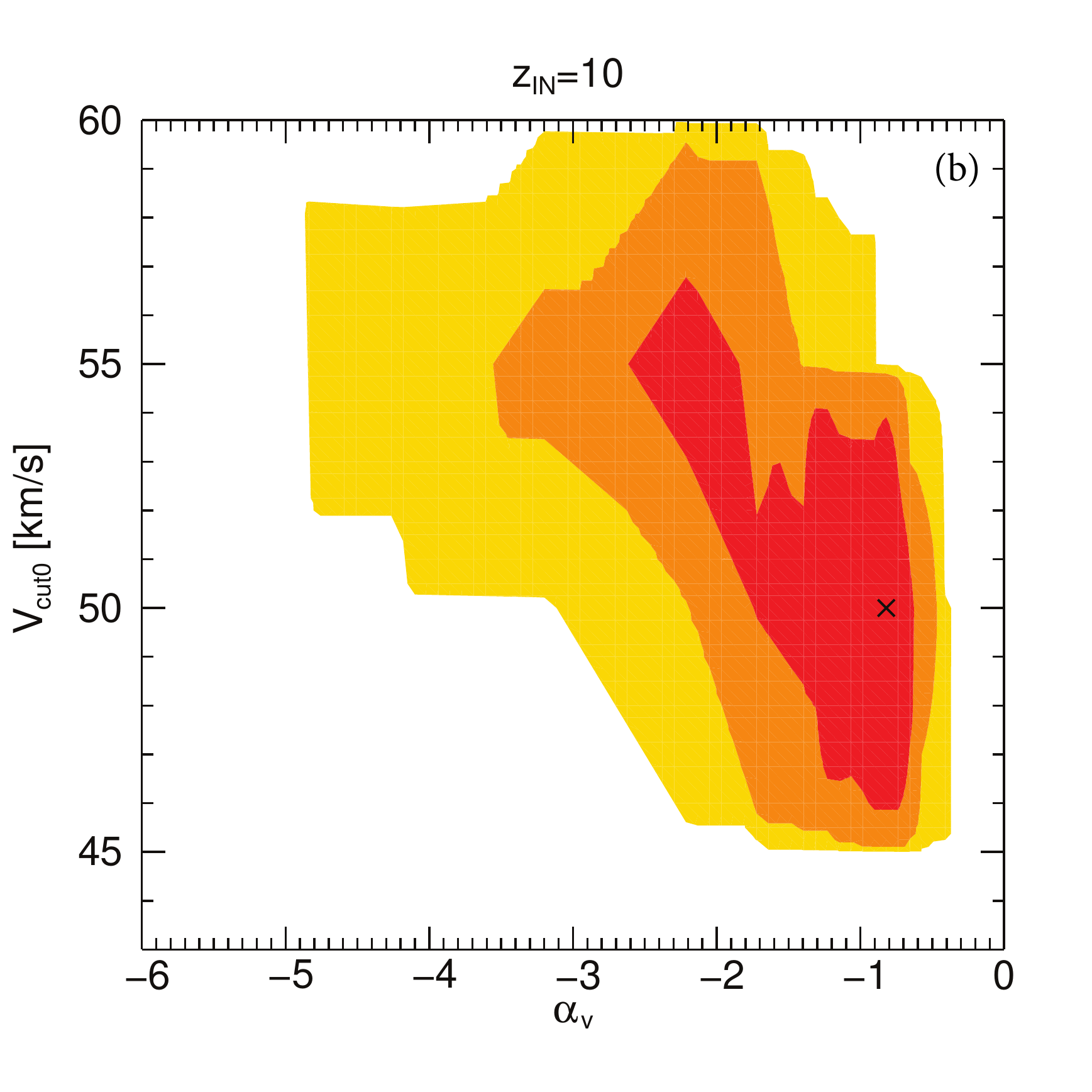}\\
\includegraphics[width=8.cm]{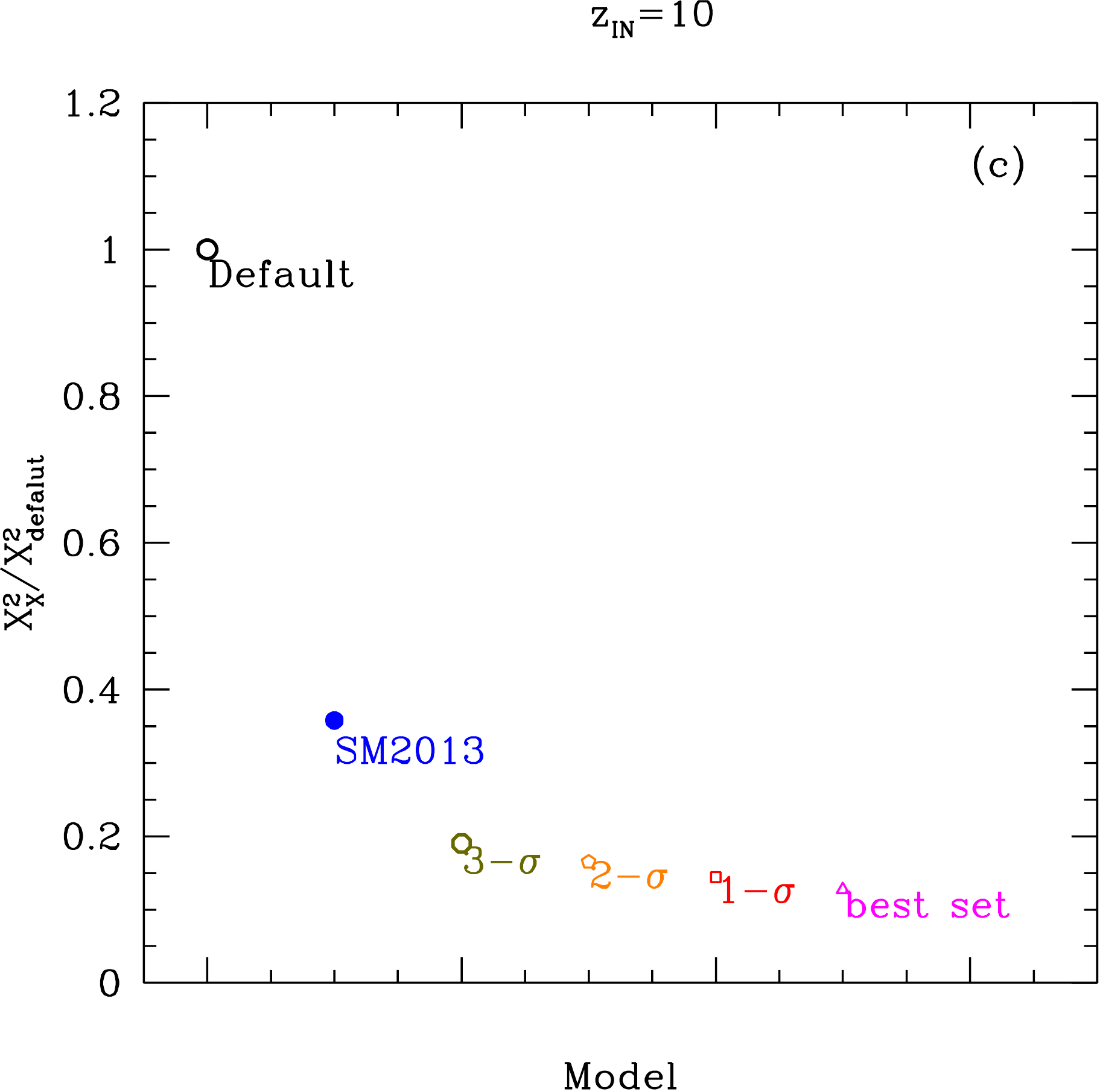}
\includegraphics[width=8.2cm]{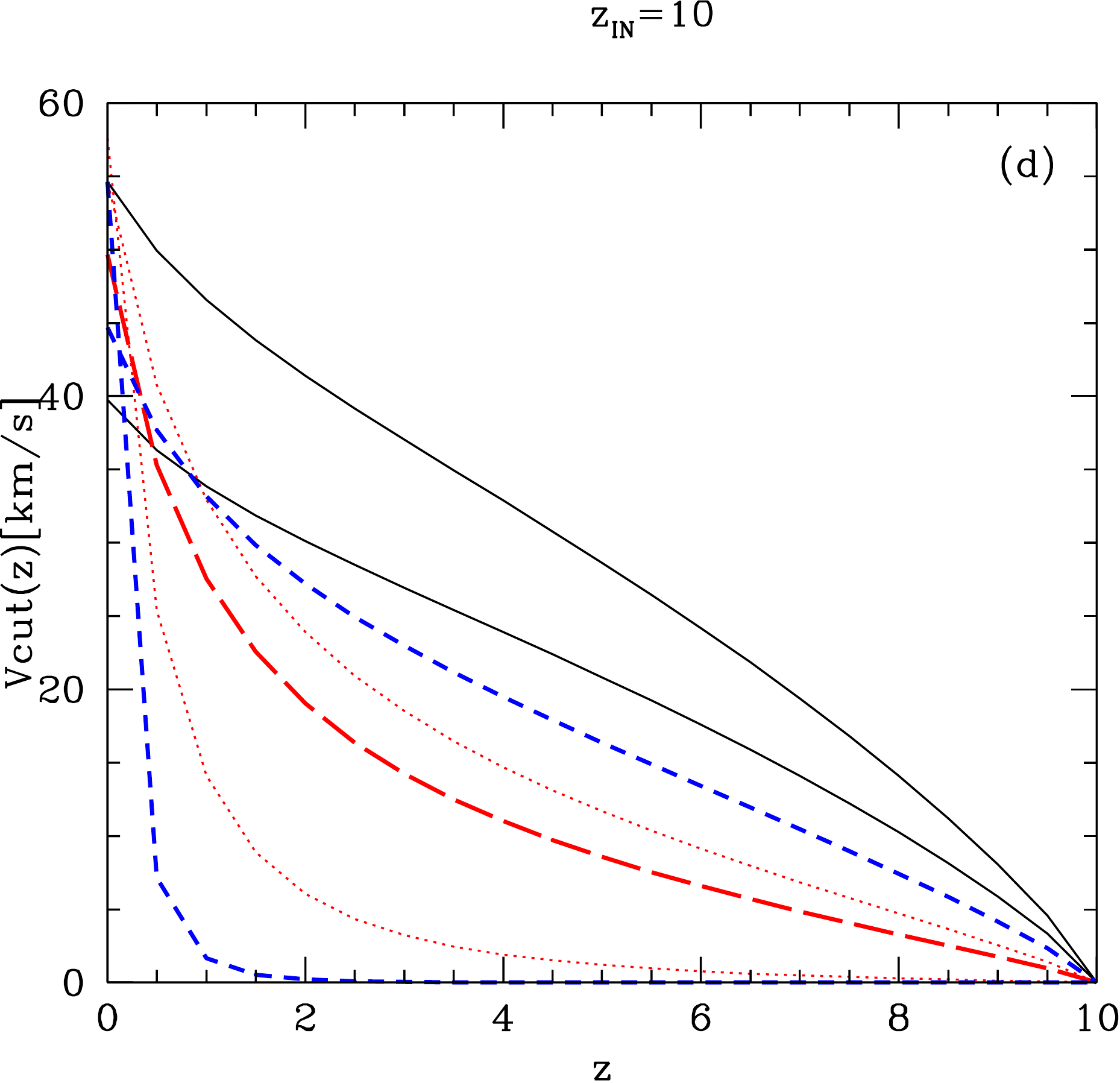}
\caption{{The upper panels (a) and (b) show the likelihood contours and the best fitting parameters $\alpha_{\rm v} - V_{\rm cut 0 }$ (indicated by the cross) for z$_{\rm IN}$=6 and z$_{\rm IN}$=10, respectively.}  The red, orange, and yellow colours indicate the 1-$\sigma$, 2-$\sigma$, and 3-$\sigma$ levels. The maximum likelihood positions are shown as a cross symbols in the figures, {$(V_{\rm cut0}, \alpha_{v})$=(55,-2.02) for z$_{IN}$=6} and $(V_{\rm cut0}, \alpha_{v})$=(50,-0.82) {for z$_{IN}$=10} . \protect \cite{sobacchi.etal.2013} suggested $\alpha_{v}$=-0.2. and $V_{\rm cut0} \sim 55 \, (40) \, {\rm km \,s}^{-1}$ {(which corresponds to the circular velocity advocated by Sobacchi \& Mesigner after reducing their critical halo mass by a factor of ten; see details in \S\ref{sec:modified})} for UV background intensities of 1 (0.01) [10$^{-21}$erg/s/Hz/cm$^{2}$/sr]. The {panel (c)} shows the ratio of $\chi^{2}$ (as defined in Eqn.~\ref{eq:chi-square}) between the default model and the model of \protect \cite{sobacchi.etal.2013} (SM2013) adopting a UVB intensity 0.1, parameters using in the three different confidence levels, and the best set of parameters in our modelling {for z$_{\rm IN}$=10}. The {panel (d)} shows the evolution of $V_{\rm cut}(z)$ for \protect \cite{sobacchi.etal.2013} (solid lines, two different UV background intensity values [0.01,1]), two examples from the 2-$\sigma$ likelihood region (dotted red lines), two examples from the 3-$\sigma$ likelihood region (dashed blue lines), and the maximum likelihood position (long dashed red line) {for z$_{\rm IN}$=10}.}
\label{Contour}
\end{figure*}

Encouraged by the success of this model in the local Universe, we next constrain the parameters of the new photo-ionisation 
model, $V_{\rm cut0}$ and $\alpha_{v}$. For combinations of values $\mbox{\boldmath $\theta$} =(V_{\rm cut0}, \alpha_{v})$, we obtain the $\chi^2$ of the model with respect to the observed HI mass function, and calculate the likelihood of each combination of parameters,
\begin{equation}
\begin{split}
    & \mathcal{L}
       \propto \exp \left(-\frac{\chi^2}{2} \right), \\
        &{\rm where}{\hspace{0.1cm}}  \chi^{2}=\sum_{i=1}^{N}\frac{[X_{i}(\mbox{\boldmath $\theta$})
        -X_{\textrm{obs},i}]^{2}}{\sigma_{\textrm{obs},i}^{2}},
    \label{eq:chi-square}
    \end{split}
\end{equation}
where $X_{i}(\mbox{\boldmath $\theta$})$  and $\sigma_{\textrm{obs},i}$ represent the model prediction for the $i^{\rm th}$ observed data point, $X_{\textrm{obs},i}$,  and its measurement error, respectively. 
From this two-dimensional likelihood distribution, we estimate the joint 1-$\sigma$ (68.3$\%$), 2-$\sigma$ (95.4$\%$), and 3-$\sigma$ (99.7$\%$) confidence levels around the best combination of parameters (those that minimize  the $\chi^{2}$). This is shown in Fig.~\ref{Contour} for $\alpha_{v}$ and $V_{\rm cut0}$. 

We find that the combination of parameters, $(V_{\rm cut0}, \alpha_{v})$, that provides the best fit to the observed HI mass function at z=0
is {$(V_{\rm cut0}, \alpha_{v})=(55 \,{\rm km \,s}^{-1},-2.02)$ for z$_{\rm IN}$=6}  and $(50 \,{\rm km \,s}^{-1},-0.82)$ {for z$_{\rm IN}$=10} (hereafter, our best fit model). {
Two values for $z_{\rm IN}$ are considered to reflect the current uncertainty in the epoch of reionization. We find that the best fitting combination of parameters for z$_{\rm IN}$=6 are in the 1-$\sigma$ confidence level of the two-dimensional likelihood distribution for the z$_{\rm IN}$=10 case, indicating that this assumption does not introduce a large uncertainty.} These values are shown in the {panels (a) and (b)} of 
Fig.~\ref{Contour} by the cross. 
The likelihoods derived from the modified model correspond to improvements in the predicted low mass end of the HI mass function, as demonstrated in the {panel (c)} of Fig.~\ref{Contour}, which shows the ratio of the $\chi^{2}$ value (defined in Eqn.~\ref{eq:chi-square}) of the default model \cite[that in][]{lagos.etal.2011b} to the value of using the Sobacchi \& Mesinger model {for z$_{\rm IN}$=10}. Also shown are the three different representative sets of parameters drawn from each confidence level in our modelling, as well as the best fitting parameter set {for z$_{\rm IN}$=10}. The {panel (d)} of Fig.~\ref{Contour} shows the evolution of $V_{\rm cut}(z)$ for different sets of parameters. The solid black lines show the evolution of $V_{\rm cut}(z)$ from \cite{sobacchi.etal.2013} for two different values of the UV background intensity\footnote{\cite{sobacchi.etal.2013} pointed out that their results are not very sensitive to the intensity of the UVB: increasing the intensity of the UVB by two orders of magnitude increases the critical mass only by a factor of 2-3. Measurements of the Lyman alpha forest imply that the UVB intensity lies between $\left(0.1 - 1\right) \times 10^{-21} \, {\rm erg \, s^{-1} \, Hz^{-1} \, cm^{-2} sr^{-1} }  $ at z=2 to 6 \cite[e.g.][]{bolton2007,HM2012}. But its evolution for $z > 6$ is very uncertain.} 
 (1 and $0.01 \times 10^{-21} {\rm erg \, s^{-1} \, Hz^{-1} \, cm^{-2} \, sr^{-1}} $.) and values for $V_{\rm cut0} \sim$  of  $55$ and $40\, {\rm km \,s}^{-1}$.
The blue dashed lines in the right panel of  Fig.~\ref{Contour} show the evolution for different combinations of parameters that lie in the 3-$\sigma$ likelihood region, while the red dotted-lines show parameter combinations that lie in the 2-$\sigma$ confidence level. The red long dashed line shows the evolution for the best fitting set of parameters {for z$_{\rm IN}$=10}. 

\begin{figure*}
\includegraphics[width=8.6cm]{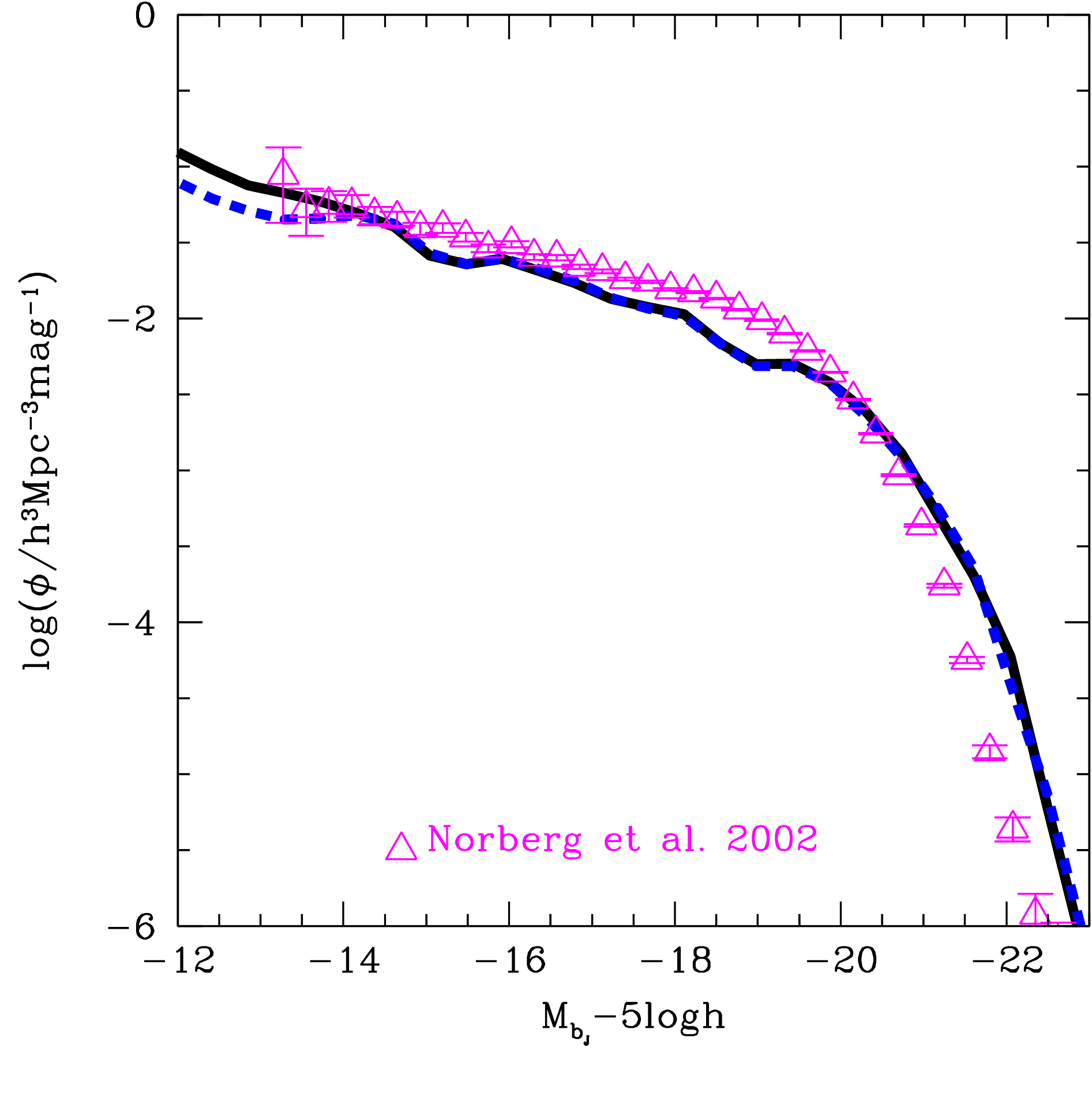}
\includegraphics[width=8.6cm]{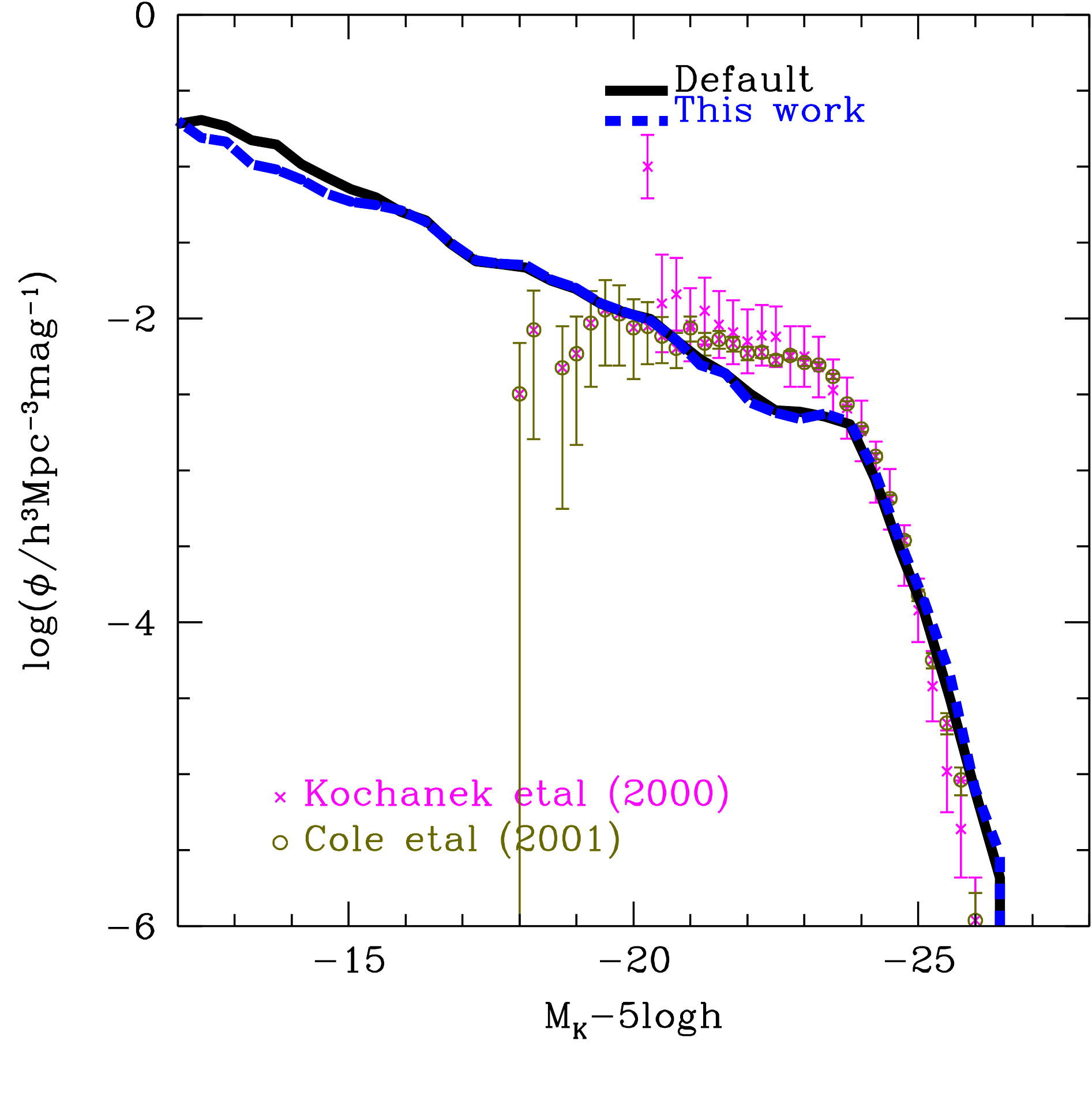}
\caption{
The b$_{J}$ band and K band galaxy luminosity function at z=0 for the default model (black solid line) and our best fit model (blue dashed line). Observations from \protect\cite{Norberg2002} for b$_{J}$ band luminosity function (left panel) and \protect\cite{cole.etal.2001} and \protect\cite{kochanek.etal.2001} for K band luminosity functions (right panel) are shown by symbols with errorbars. 
}\label{EX}
\end{figure*}

We note that in this exercise, ${\alpha_{v}}$ and $V_{\rm cut0}$ are varied independently of the other parameters in GALFORM. We have verified that the modified model for photoionisation feedback using these parameters does not affect other model predictions that were in good agreement with the observations. This is because these observations tend to probe more massive galaxies than the low-mass end of the HI mass function. For example, Fig.~\ref{EX} shows the predicted b$_{\rm J}$ and K-band luminosity functions in the default model and in the new model, demonstrating that we find broad consistency with observations for galaxies brighter than -15 mag.

\subsection{The distribution of HI in central and satellite galaxies}

\begin{figure*}
\includegraphics[width=8.6cm]{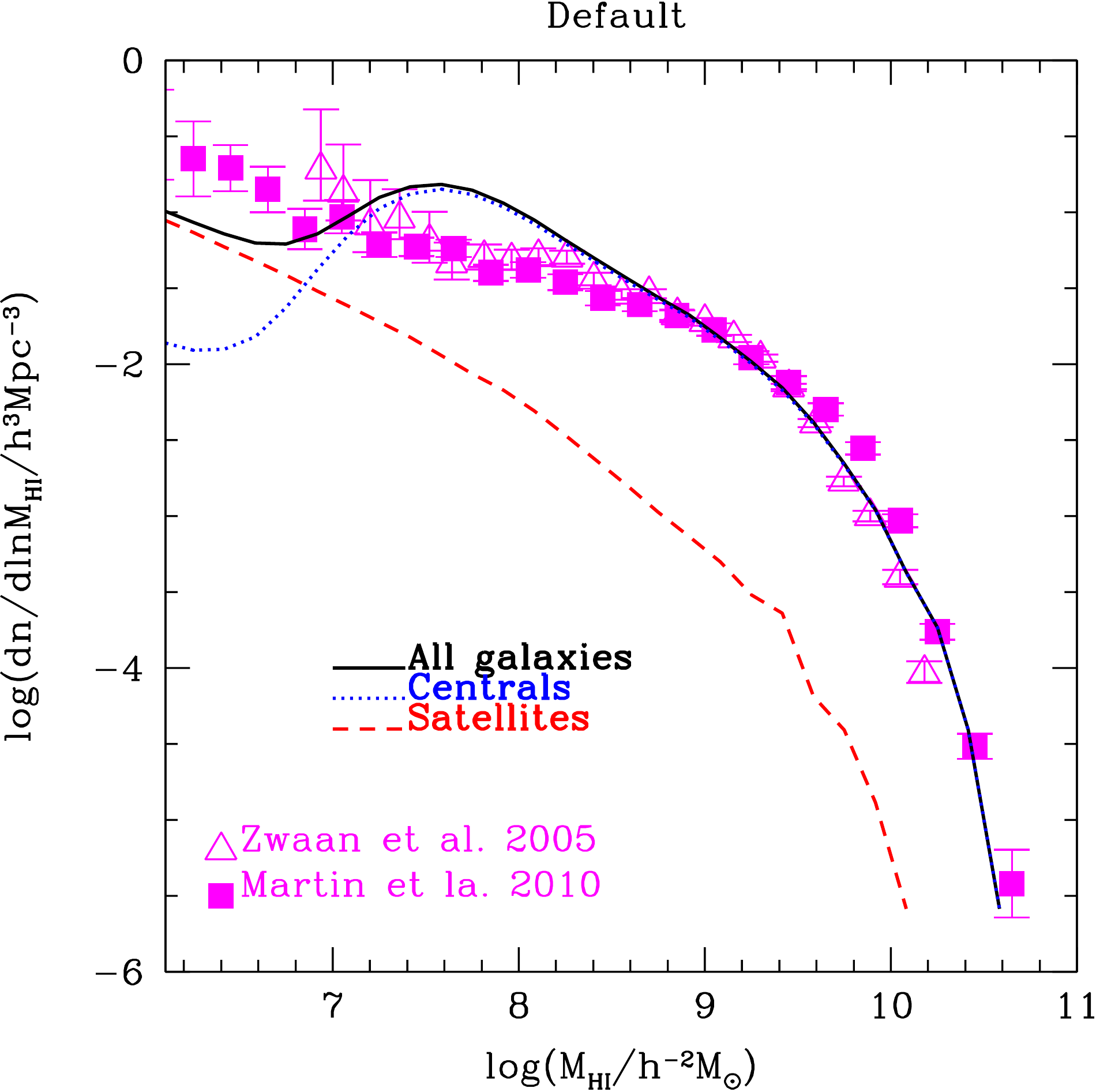}
\includegraphics[width=8.6cm]{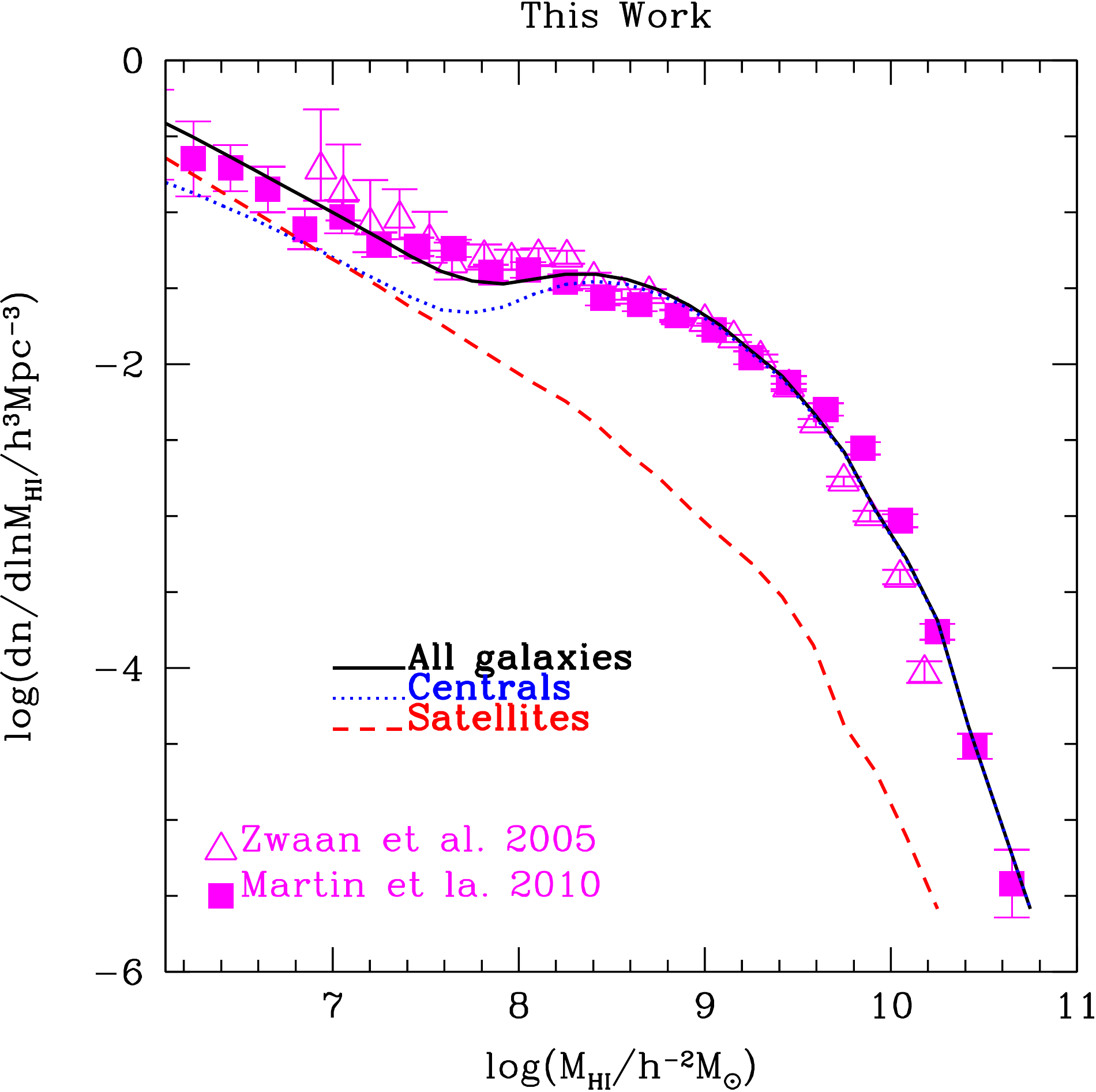}
\caption{The HI mass functions at z=0 (black solid lines) in the default model (left panel) and the best fit in our new model (right panel). The contributions to the HI mass function, from central galaxies is shown by blue dotted lines and from satellite galaxies as red dashed lines. 
}\label{HIMFCS}
\end{figure*} 

In Fig.~\ref{HIMFCS} we investigate the contributions of satellite and central galaxies to the HI mass function in the default model and in our best fit model. {The non-monotonic feature seen in the HI mass function from the default model is related to both the suppression of cooling process in central galaxies, and the abundance of satellite galaxies at low HI masses. Photoionisation feedback in the default model suppresses the cooling of gas in central galaxies which have circular velocities less than $V_{\rm cut}$. As a result, the contribution of central galaxies to the number density of galaxies with $M_{\rm HI}\sim 10^{6}h^{-2}M_{\odot}$ is less than 10\% in the default model. On the other hand, satellite galaxies in massive dark matter haloes ($M_{\rm halo}> 10^{13}h^{-1}M_{\odot}$) contribute more than 90\% of the abundance at galaxy masses of $M_{\rm HI}\sim10^{6}h^{-2}M_{\odot}$. Furthermore, in the default model the abundance of satellite galaxies monotonically increases towards low HI mass. The reason why the HIMF of satellite galaxies does not show the dip at low HI masses observed in the HIMF of central galaxies is that the satellites were mainly formed before reionization. Thus the sum of the abundance of central galaxies and satellite galaxies introduces a non-monotonic feature in the default model around the $M_{\rm HI}$ that is directly related to the value of the parameter $V_{\rm cut}$.}

The contribution of satellite galaxies to the HI mass function in our best fit model is larger than in the default model across the whole HI mass range shown. Central galaxies dominate the abundance of galaxies at $M_{\rm HI} \ge 10^{8}\,h ^{-2}\, {\rm M_{\odot}}$ in both models. However, at $M_{\rm HI} \le {10^{7}}\, h ^{-2}\,{\rm M_{\odot}}$, the abundance of central galaxies drops below the abundance of satellite galaxies in the default model. In contrast, our best fit model includes a similar abundance of central and satellite galaxies. The reason is that central galaxies with circular velocities from $30 \, {\rm km \, s}^{-1}$ to $ 50 \, {\rm km \,s}^{-1}$ in the default model are never affected by photoionisation feedback, while in the new model these are suppressed between 0$\le$$z$$\le$1. These central galaxies dominate the HI mass function at 
$10^{6}\,h^{-2}\,{\rm M_{\odot}} \le M_{\rm HI} \le 10^{8} h ^{-2}\, {\rm M_{\odot}}$. 
In addition, central galaxies with circular velocities $\le \, 30 \, {\rm km \,s}^{-1}$ in the default model cannot contain cold gas at $z$$\le$10, and cease forming stars shortly after. In the new model these galaxies are able to continue accreting new gas and form stars down to $z$$\sim$1. This produces the increased abundance of central galaxies at $M_{\rm HI} \le 10^{7} \, h^{-2} \, {\rm M_{\odot}}$ that reside in haloes of mass $\le 10^{10} \,h^{-1} {\rm M_{\odot}}$.

{This difference in the abundance of satellite versus central HI galaxies explains why the non-monotonic feature seen in the default model is not found in our new model, providing evidence for an evolution in the strength of photoionisation feedback.}

\section{Other probes of photoionisation feedback}
In this section, we examine the effects of the modified model for photoionisation feedback on the predicted spatial distribution of galaxies selected by their HI mass, and on the relation between the HI and the stellar mass contents of galaxies at z=0. 

\subsection{Clustering of HI galaxies}
We first use the two-point correlation function to investigate how the modified photoionisation feedback influences the spatial distribution
of galaxies selected by their HI mass. The distribution of HI-selected galaxies in the default model agrees well with the observations of \cite{Meyer2007} and \cite{martin.etal.2010} who used the HIPASS and the ALFALFA surveys, respectively, to measure the clustering of galaxies with $M_{\rm HI} > 10 ^{8} h^{-2} {\rm M_{\odot}}$ \citep{Kim2013}. However, it is challenging to measure the spatial distribution of HI-poor galaxies due to the small number of such galaxies observed in current surveys \citep[cf.][]{Papa2013}. 

\begin{figure}\includegraphics[width=8.5cm]{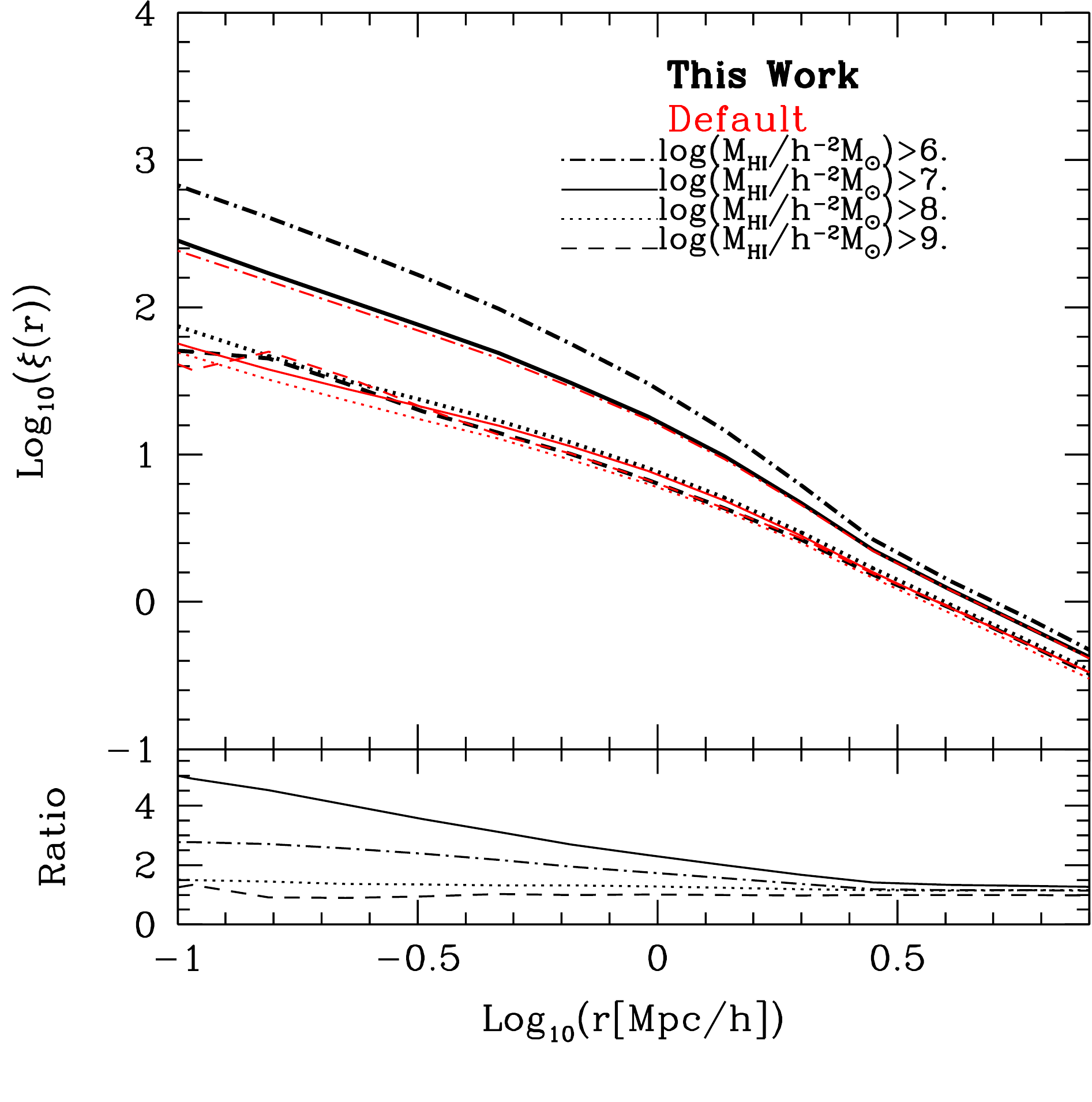}
\caption{The impact of the modified photoionization feedback model on the two-point correlation function. The  lines show the predictions for four galaxy samples defined by different HI mass thresholds (see key). The default model (our best fitting model) is shown by the thin red (thick black) lines. The lower subpanel shows the ratio of the prediction from the best fitting model to that from the default model for different HI mass thresholds.
}\label{2DCF}
\end{figure}

The top panel of Fig.~\ref{2DCF} shows the predicted two-point correlation functions measured from the default model and for our best fitting model, while the bottom panel shows the ratio between the correlation function in the best fit model to that in the default model. We consider four HI mass selections in order to see how the modified model for photoionisation feedback is imprinted on the clustering of HI-selected galaxies. The predicted two-point correlation functions show that the clustering amplitude in the default model is much lower than in our best fitting model for the HI mass samples with $M_{\rm HI} > 10^{6} h^{-2} {\rm M_{\odot}}$ and $M_{\rm HI} > 10^{7} h^{-2} {\rm M_{\odot}}$. In addition, the slope of the two-point correlation function at small separations is steeper in our best fitting model than it is in the default model. The ratios of correlation functions for the HI mass samples with $M_{\rm HI} > 10^{6} h^{-2} {\rm M_{\odot}}$ and $M_{\rm HI} > 10^{7} h^{-2} {\rm M_{\odot}}$ shows a large difference at small separations. This is because the two models have different numbers of satellite galaxies with HI masses in the range $10^{6} h^{-2} {\rm M_{\odot}}$ and $10^{8} h^{-2} {\rm M_{\odot}}$, which dominate the small scale clustering signal. 

The increase in the observed number of galaxies with $M_{\rm HI} < 10^{7} h^{-2} {\rm M_{\odot}}$ expected from ongoing and future HI galaxy surveys (eg. ASKAP, MeerKAT, SKA) promises to allow the nature of photoionisation feedback to be inferred from the spatial distribution of galaxies with low HI mass.   

\subsection{Stellar mass -- HI mass relation}
\begin{figure}
\includegraphics[width=8.5cm]{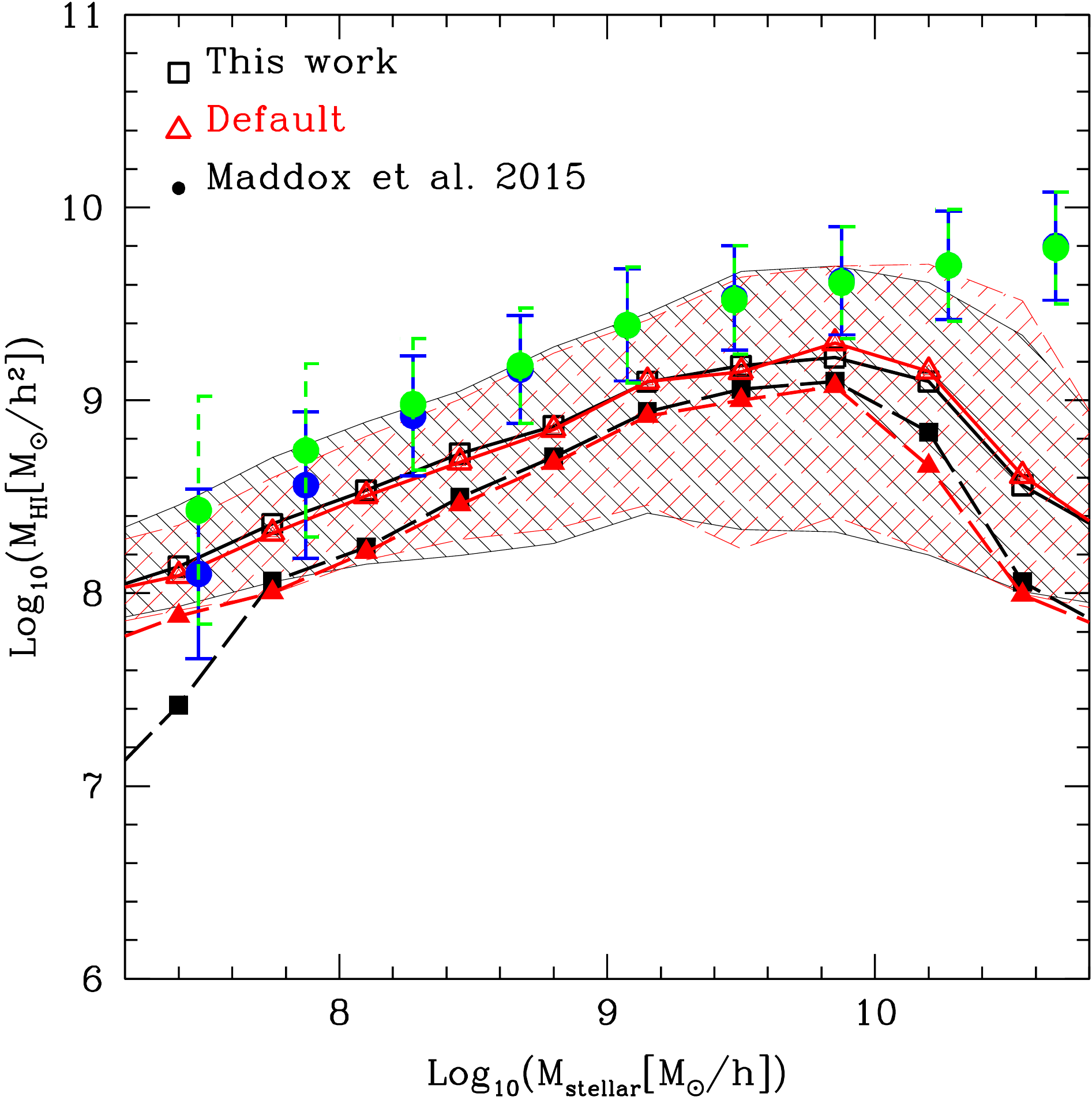}
\caption{The HI mass as a function of the stellar mass for the default model and our best fit model with only galaxies which have HI mass $> 10^{8} h^{-2} {\rm M_{\odot}}$, stellar mass $> 10^{7} h^{-1} {\rm M_{\odot}}$ and the bulge-to-total stellar mass ratio $<$ 0.5. The median HI masses in bins of stellar mass are shown for the default model as red empty triangles and our best fit model as black empty squares. The shaded regions represent the 16\% and 84\% range for the default model (red dashed) and our best fit model ( black solid). The filled circles represent the observations in \protect\cite{Maddox2015}. Blue points are derived only from galaxies with SDSS spectra, and green points are derived only from galaxies without spectrum. The error bars correspond to the 1-$\sigma$ dispersion of observation. In comparison, the median HI masses in bins of stellar mass are shown for the default model as red filled triangles and our best fit model as black filled squares using lower HI mass threshold (M$_{\rm HI}$$> 10^{6} h^{-2} {\rm M_{\odot}}$).
}\label{DISMHIMS}
\end{figure}

Fig.~\ref{DISMHIMS} shows predictions for the relation between the HI and stellar mass contents of galaxies for both the default model and our best fit model. These are compared with the observational results for the median HI mass as a function of stellar mass  presented by \cite{Maddox2015} (filled circles). Note that we only consider model galaxies which have a HI mass $> 10^{8} h^{-2} {\rm M_{\odot}}$, a stellar mass $> 10^{7} h^{-1} {\rm M_{\odot}}$ and a bulge-to-total stellar mass ratio $<0.5$ in this plot, in order to approximately imitate the selection criteria of the ALFALFA (N. Maddox, priv. communication). As argued by \cite{Maddox2015}, this relation provides a fundamental benchmark for galaxy formation models by combining HI mass data with stellar mass data. The relation between the HI mass and stellar mass in both models remains in reasonable agreement with the 1-$\sigma$ dispersion of the observations. We note that high stellar mass galaxies in the models typically lie slightly below the observations. This could be due in part to remaining differences between the observational sample selection and the crude cuts applied to the model galaxies to mimic that described above. Indeed, HI observations of stellar-mass selected samples show somewhat different behaviour and are in better agreement with the model predictions \cite[]{Catinella2010, Serra2012}. Further analysis of the HI gas fraction distribution function of galaxies at $z$$\sim$0 from GALFORM and comparison with observations are described in \cite{Lagos2014}.

Finally, we note that the model predictions shown in Fig.~\ref{DISMHIMS} are the true stellar masses rather than those estimated from photometry or spectra. \cite{Mitchell2013} showed that the process of estimating the stellar mass of a galaxy can lead to differences from the stellar mass predicted directly by the model. These include systematic shifts,  which arise from a mismatch between the choice of stellar initial mass function used to estimate the stellar mass and that adopted in the semi-analytic model and from differences in the modelling of dust extinction, and scatter, due to variation in the assumed metallicity and star formation history. If we replaced the true model stellar masses in Fig.~\ref{DISMHIMS} with estimated stellar masses, this would soften the downturn in the relation at high stellar masses.   
We also include the predictions from the two models using a lower HI mass threshold (M$_{\rm HI}$$>$10$^{6}h^{-2}$M$_{\odot}$ instead of M$_{\rm HI}$$>$10$^{8}h^{-2}$M$_{\odot}$; see red filled triangles for the default model and the black filled squares for our best fitting new model connected by long dashed lines for the median HI masses in bins of stellar mass). The predictions deviate from one another at stellar masses between $10^{7} h^{-1} {\rm M_{\odot}}$ and $10^{8} h^{-1} {\rm M_{\odot}}$. This stellar mass range corresponds to HI masses between $10^{6} h^{-2} {\rm M_{\odot}}$ and $10^{8} h^{-2} {\rm M_{\odot}}$, where the two models also predict different HI abundances because of the different modelling of photoionisation feedback. 

\subsection{Evolution of HI density}
\begin{figure}
\includegraphics[width=8.8cm]{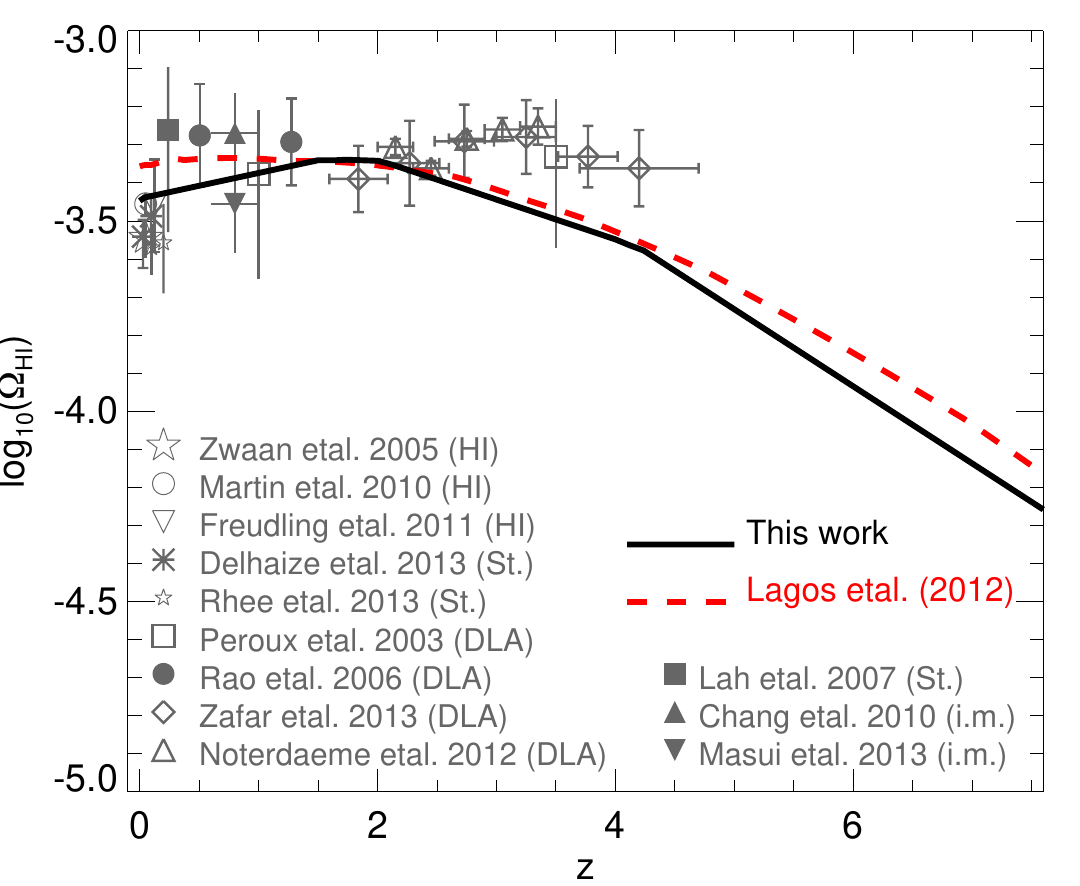}
\caption{The HI density in units of the critical density as a function of redshift for our $V_{\rm cut}(z)$ model (this work, black solid line) and \citet{lagos.etal.2011b} model (red dashed line). The points show various observations. The parenthesis after each observation indicated the method used are HI: direct 21cm line detection \citep{Zwaan05,martin.etal.2010,Freudling11}, St.: stacking of 21cm line \citep{Lah07,Delhaize13,Rhee13}, DLA: absorption line studies \citep{Peroux03,Rao06,Noterdaeme12,Zafar13}, and i.m.: intensity mapping of the 21cm line \citep{Chang10,Masui13}. 
}\label{Omega}
\end{figure}

Finally we investigate the impact of our new model on the HI density as a function of redshift. Fig.~\ref{Omega} shows $\Omega_{\rm HI}$, defined as the ratio between the mean HI density and the critical density of the Universe, for the new evolving $V_{\rm cut}(z)$ model (black solid line) and the \citet{lagos.etal.2011b} model (red dashed line, the default model).  The HI density predicted using the new model with $V_{\rm cut}(z)$ shows slightly better agreement with the observed HI density at z$<$0.1 than the \citet{lagos.etal.2011b} model, recovering the decreases $\Omega_{\rm HI}$ suggested by the observations of $z$$\ge$0.2. However, the difference between the two models are within the uncertainties of the observations, and new constraints at intermediate redshifts are needed.

\section{Summary}
\label{sec:summary}

One of the most important ingredients in a model of galaxy formation is the description of neutral hydrogen 
which provides the fuel for star formation. The HIPASS \cite[]{Zwaan05} and ALFALFA \cite[]{martin.etal.2010} surveys have been used to estimate the HI mass function down to masses of a few times $10^{6} \,h^{-2}\, {\rm M_{\odot}}$ in the local Universe. Attempting to reproduce these observed HI mass functions using a galaxy formation model allows us to probe the physics of low mass galaxy formation.

In the standard implementation of GALFORM the effect of photoionisation feedback is modelled by restricting galaxy formation in halos which correspond to effective circular velocities smaller that a fixed circular velocity of $V_{\rm cut}$ at redshifts following the end of reionization $z_{\rm cut}$ (the default model). Halos with circular velocities below $V_{\rm cut}$ are not allowed to cool gas at redshifts below $z_{\rm cut}$. The predicted HI mass function from this default model shows good agreement with observations of the HI mass function for masses greater than $M_{\rm HI} \sim 10^{8} \, h^{-2} \, {\rm M_{\odot}}$. However, the default model fails to explain the abundance of galaxies with HI masses between $10^{6} h^{-2} {\rm M_{\odot}}$ and $10^{8} \,h^{-2}\, {\rm M_{\odot}}$. 

There are several possibilities to ease the discrepancy at the low mass end of the HI mass function between observations and the default model, including chaning the supernova feedback, cosmological parameters and the form of the star formation law in the model. In particular \cite{kim2013a} showed that photoionisation feedback is an important physical processes for the low mass end of HI mass function. We have investigated these possibilities using a version of the GALFORM semi-analytical galaxy formation model \citep[cf.][]{cole.etal.2000} by \citet{lagos.etal.2011b} to simulate the low mass end of the HI mass function. In our investigation we showed that varying the SNe feedback, the efficiency of star formation and the cosmological model all failed to explain the shape and abundance of the low mass end of HI mass function. Varying the redshift and critical velocity of galaxy suppression of formation by photoionisation feedback from reionization is also unable to explain the shape of the low mass end of HI mass function. 

To understand what photoionisation physics affects the shape and abundance of the HI mass function in the local Universe, we introduced a redshift dependent photoionisation feedback into GALFORM, motivated by the simulations of \cite{sobacchi.etal.2013}. This illustrates the modular nature of semi-analytical modellling, whereby the treatment of a given physical process can be overhauled and improved when new information becomes available. We find that redshift dependent modelling of feedback from photoionisation on low mass galaxy formation is needed to explain the shape and abundance of the observed HI mass function in the local Universe. We also find that the sensitivity of the HI mass function to the redshift evolution of photoionisation feedback strength is larger at high redshift. Our modelling suggests that future measurements of HI clustering in low mass galaxies {and the relation between the HI mass and the stellar mass of galaxies} will provide additional constraints on the form of ionising feedback.

Ongoing and future HI selected galaxy surveys on the SKA and its pathfinders, such as ASKAP, MeerKAT are expected to extend our view of the HI Universe to higher redshifts not only for the HI mass function but also for the distribution of HI galaxies. The evolution of the HI mass function as a function of redshift will tell  us about the physical processes that drive low mass galaxy formation, including photoionisation feedback. 

\section*{Acknowledgements}

H-SK is supported by a Discovery Early Career Researcher Awards from the Australian Research 
Council (DE140100940). CP thanks Simon Driver and Aaron Robotham for helpful discussions. CP is supported by DP130100117, DP140100198, and FT130100041. CL is funded by the ARC project DE150100618.
This work was supported by a STFC rolling grant at Durham. The 
calculations for this paper were performed on the ICC Cosmology Machine, 
which is part of the DiRAC Facility jointly funded by the STFC, the Large 
Facilities Capital Fund of BIS, and Durham University.  Part of the research 
presented in this paper was undertaken as part of the Survey Simulation 
Pipeline (SSimPL; {\texttt{http://www.astronomy.swin.edu.au/SSimPL/}). The 
Centre for All-Sky Astrophysics is an Australian Research Council Centre of 
Excellence, funded by grant CE11E0090. 

\newcommand{\noopsort}[1]{}

\bibliographystyle{mn2e}

\bibliography{HIMF}

\label{lastpage}
\end{document}